\pdfoutput=1
\documentclass[pdflatex]{sn-jnl}

\usepackage{float}
\usepackage{circuitikz}
\usepackage{natbib}
\usepackage{booktabs}       
\usepackage{array}          
\usepackage{amsfonts}       
\usepackage{graphicx}
\graphicspath{{images/}}
\usepackage{subcaption}
\usepackage{amsmath}
\usepackage{tikz}
\usepackage{quantikz}

\author*[1]{\fnm{William} \sur{Cashman}}\email{wcashman1351@gmail.com}

\author[2]{\fnm{Marco} \sur{Caselli}}\email{marco.caselli@and-e.com}

\author[2]{\fnm{Saad} \sur{Hamid}}\email{saad.hamid@and-e.com}

\author[3]{\fnm{Daniel} \sur{Mills}}\email{daniel.mills@quantinuum.com}

\author[4]{\fnm{Pavel} \sur{Stishenko}}\email{pstishenko@gmail.com}

\affil*[1]{\orgdiv{Department of Computer Science}, \orgname{University of Oxford}, \orgaddress{\street{15 Parks Rd}, \city{Oxford}, \postcode{OX1 3QD}, \country{United Kingdom}}}

\affil[2]{\orgname{OXFIDA}, \orgaddress{\street{R3, 46 Woodstock Road}, \city{Oxford}, \postcode{OX2 6HT}, \country{United Kingdom}}}

\affil[3]{\orgname{Quantinuum}, \orgaddress{\street{13–15 Hills Road}, \city{Cambridge}, \postcode{CB2 1NL}, \country{United Kingdom}}}

\affil[4]{\orgdiv{Cardiff Catalysis Institute, School of Chemistry}, \orgname{Cardiff University}, \orgaddress{\street{Park Place}, \city{Cardiff}, \postcode{CF10 3AT}, \country{United Kingdom}}}

\begin{document}

\title{Insurance portfolio simulation with a quantum circuit}

\abstract{A central challenge in actuarial science is characterising the distribution of losses generated by an insurance portfolio.
Copulas provide a natural framework for modelling joint distributions, but flexible high-dimensional copula fitting and sampling remains challenging in practice.
We propose a Quantum Circuit Born Machine (QCBM) approach adapted to insurance portfolios, combining (i) a tail-focused discretisation of heavy-tailed marginals, (ii) a Monte Carlo-style objective function that enables the numerical precision of the algorithm to scale when trained on a relatively small number of samples, and (iii) a two-stage hierarchical training scheme enabling efficient training on a simulator and fine-tuning on quantum hardware.
On a mixed portfolio of equity returns and insurance losses, the trained circuit recovers the qualitative dependence structure, with near-independence between equity and losses and positive dependence among loss lines. Exponential quantile encoding substantially improves the fit to the tail of the aggregate loss distribution relative to a uniform encoding, and the hierarchical scheme substantially outperforms single-stage training of a circuit of the same size from random initialisation.}

\maketitle

\section{Introduction}

\subsection{Insurance portfolio risk modelling}

The essence of any insurance business is trading the risk of losses due to adverse events for a fixed insurance premium. Insurers benefit from portfolio diversification: aggregating many weakly correlated risks reduces the variance of total portfolio losses. In practice, however, individual risks are rarely independent, and dependence is particularly important in the tails, where common drivers can produce simultaneous large losses across multiple lines of business. For example, home and motor insurance may exhibit modest average correlation while still sharing substantial upper-tail dependence during severe weather events such as hurricanes, floods, or hailstorms.

Insurers also invest premium income and hold sizeable asset portfolios on their balance sheet. Regulators impose capital requirements calibrated to tail-sensitive risk measures such as tail value-at-risk. Large natural disasters can therefore affect both sides of the balance sheet: they generate insurance claims while simultaneously depressing asset values \cite{bordeau-brien_impact_2017,sakariyahu_natural_2023,zhao_drought_2025}. Modelling the distribution of total portfolio outcomes consequently requires a joint model over both liabilities and assets.
Given such a model, we can sample from the joint distribution to estimate aggregate losses and compute portfolio-level risk measures. These quantities then inform pricing, capital allocation, and reserving decisions.

In this paper, we study the problem of fitting a joint distribution to empirical data from a mixed portfolio of insurance losses and equity returns, with particular attention paid to techniques that improve the fit in the upper tails of the loss distributions --- a well-known challenge for generative models of heavy-tailed data with limited samples.

\subsection{Copula-based models}

The most computationally challenging part of fitting a joint distribution is learning the dependence structure between its variables. A common approach is to separate the marginal distributions of the variables from their joint dependence; the latter, once expressed on uniform marginals, is called a \textit{copula}. Since \citeauthor{Sklar1959}'s foundational work~\citep{Sklar1959}, copulas have become a central tool in financial and actuarial risk modelling. Standard references include \citet{McNeil2015} and the survey by \citet{pfeifer2003modeling}.

Sklar's Theorem states that any multidimensional joint distribution function $F(x_1, x_2, \ldots, x_d)$ can be decomposed as follows:

\begin{equation}
  \label{eq:Sklar}
  F(x_1, x_2, \ldots, x_d) = C(F_1(x_1), F_2(x_2), \ldots, F_d(x_d))
\end{equation}

Here $F_i(x_i)$ denotes the marginal distribution function $F_i(x_i) \equiv \Pr(X_i \leq x_i)$, and $C$ is a copula $C: [0,1]^d \rightarrow [0,1]$. Marginals are often comparatively easy to fit because there are many well-studied one-dimensional distribution families. By contrast, fitting and sampling a flexible copula becomes difficult as the dimension grows. Equation~\eqref{eq:Sklar} is therefore useful because it separates marginal fitting from dependence modelling. After fitting the marginals, we transform each sample as

\begin{equation}
\label{eq:uniformization}
u_i \equiv F_i(x_i)
\end{equation}

Each $u_i$ then has a uniform marginal distribution on $[0,1]$, and the joint distribution of $\{u_i\}$ is precisely the copula $C$. This preprocessing isolates the task of fitting the joint dependence structure independently of the marginal distributions, while allowing samples to be mapped back to the original scale via the inverse marginals $x_i = F_i^{-1}(u_i)$.

Common copula models include parametric families such as the Gaussian and Archimedean copulas, and vine copulas~\citep{bedford2002vines,aas2009pair}, which build a multivariate dependence structure from bivariate (pair) copulas arranged on a tree hierarchy. Their development is driven by three competing requirements: expressive power, tractable fitting, and efficient sampling. Satisfying all three simultaneously is difficult, which motivates application-specific copula constructions.

Recent machine-learning approaches offer a promising route to flexible copula modelling. In particular, generative models and their quantum analogues have been used to fit portfolio copulas \cite{zhu2022generative,zhu2023copula}. A key appeal of quantum circuits in this setting is that a circuit on $n$ qubits defines a distribution over $2^n$ outcomes with a compact parameterisation.

\subsection{Approximating copulas with quantum circuits}

We build on \citet{zhu2023copula}, who demonstrated that Quantum Circuit Born Machines (QCBMs) can be trained to learn copula distributions. In that setting, a parameterised circuit is optimised so that, when applied to the ground state $|0\rangle^{\otimes n}$, measurement outcomes in the computational basis follow the target copula, so each shot returns an independent draw from the model. In our approach, the model distribution is instead defined on the amplitude magnitudes of the circuit state (Section~\ref{sec:amplitude-objective}). On hardware it is obtained by estimating the measurement distribution from shots and applying a square-root renormalisation, after which samples are drawn classically from the result.

Our contribution is to adapt this framework to mixed portfolios of equity returns and insurance losses, where heavy tails and limited training data make standard QCBM training unstable. The resulting method combines (i) a tail-focused discretisation of heavy-tailed marginals, (ii) a randomised KL-divergence objective that provides a stable training signal even when samples are sparse relative to the discretised support, which is what enables the numerical precision of the algorithm to scale, and (iii) a two-stage training scheme, building on the hierarchical learning approach of \citet{gharibyan2023hierarchical}, that warm-starts a large circuit on quantum hardware from a smaller one pretrained on a statevector simulator.

Section~\ref{sec:methods} develops the encoding, objective, and hierarchical training strategy. Section~\ref{sec:training-pipeline} describes the full two-stage training pipeline that combines these components, Section~\ref{sec:results} reports experimental results on an obfuscated insurance-loss dataset and a synthetic Gumbel benchmark, and Section~\ref{sec:conclusions} discusses the findings and their limitations.

\section{Methods}
\label{sec:methods}

\subsection{Encoding stock returns and insurance losses}

To model insurance losses with a QCBM~\citep{benedetti2019generative,liu2018differentiable}, the data must first be discretised and encoded into computational basis states. We follow the standard copula workflow of fitting marginals and transforming observations into approximately uniform variables, with additional modifications chosen to improve training stability.

Our dataset contains both stock returns and insurance losses. For stock returns, we first transform each variable to have a uniform marginal distribution on $[0,1]$. We then discretise that interval uniformly according to the number of states available to each variable in our circuit. With $n$ qubits this uniform encoding assigns each interval of width $2^{-n}$ to a different basis state. It is appropriate for distributions without heavy tails, as observed in \cite{zhu2023copula}.

However, insurance losses have heavy tails, and a uniform discretisation allocates too little resolution to the most extreme outcomes when only a small number of qubits is available.  We therefore use an exponential quantile encoding that assigns progressively finer resolution to higher quantiles.

This encoding scheme assigns the first 50\% of observations to the first bin, the second bin contains the next 25\%, the third bin the next 12.5\%, and so on. The final bin contains the remaining mass. 
A model with $n$ qubits per variable therefore has $2^n$ bins, with the final bin containing the top $2^{1-2^{n}}$ quantile of values. 
In the datasets studied here, three qubits per variable gives a top quantile of $2^{-7} = 1/128$, which is already sufficient to preserve useful tail information.

Decoding proceeds by sampling uniformly from the selected bin in the original variable's domain.
Training is carried out incrementally, beginning with a smaller model and then transferring the learned parameters to a larger model with more qubits per variable.
To make this transfer step consistent with the encoding, each added qubit introduces a further uniform split of the previous bins, effectively doubling the precision of the encoding.

\subsection{Randomised KL-divergence}

The Kullback--Leibler (KL) divergence measures how much one probability distribution diverges from another. For discrete distributions $P$ and $Q$ on the same sample space $\chi$, it is defined as

\[
D_{KL}(P, Q) := \sum_{x \in \chi} P(x) \ln \left( \frac{P(x)}{Q(x)} \right)
\]

$D_{KL}$ is not symmetric, and is infinite whenever there exists $x \in \chi$ such that $P(x) > 0$ but $Q(x) = 0$. Because we are training our model on relatively few samples compared to the size of the sample space, we use a clipped form,

\[
D^\epsilon_{KL}(P, Q) := \sum_{x \in \chi} P(x) \ln \left( \frac{\max\{\epsilon, P(x)\}}{\max\{\epsilon, Q(x)\}} \right)
\]

with $\epsilon = 10^{-6}$ in all experiments.

\subsubsection{Problems with KL-divergence}

The numerical estimate of the KL divergence becomes unreliable when there are few observations relative to the size of the sample space. With four variables and three qubits per variable, the encoded sample space already contains $(2^3)^4 = 4096$ states, and with four qubits per variable it contains 65{,}536. Our Stock/Insurance dataset, in contrast, has 939 weekly observations, so most states are never observed. The gap widens quickly as variables or qubits are added.

To obtain a stable estimate of the divergence between model and empirical distributions when the number of samples is small, we have adopted a randomised approach to divergence estimation.

\subsubsection{Construction of randomised KL divergence}

Inspired by the successful application of Monte Carlo techniques to the estimation of multidimensional integrals, we propose the following algorithm to estimate the divergence.

\begin{enumerate}
    \item For each dimension, choose a random boundary between discretisation bins and merge the bins on either side to form two larger bins, which may have different sizes.
    \item Compute the clipped KL divergence on that randomised aggregation.
    \item Repeat the procedure and average the resulting divergences. We use 10 randomisations. This value was chosen empirically, because it gave a better fit in the upper tail than 100 randomisations in our experiments; we did not perform a systematic sweep.
\end{enumerate}

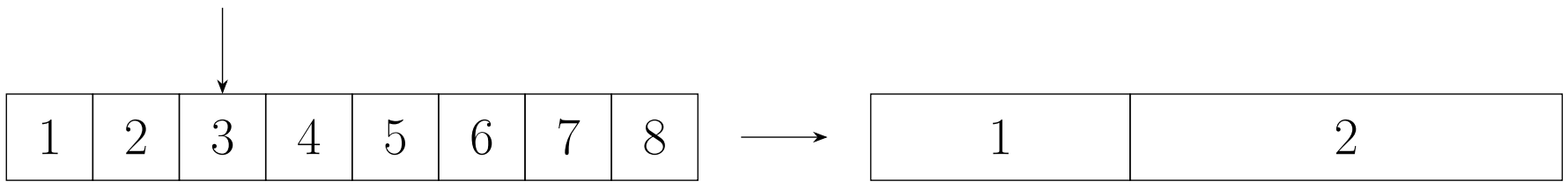
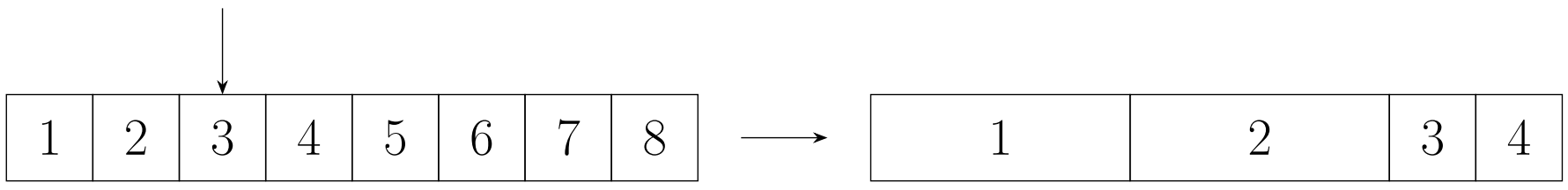
\begin{figure}
\centering
\begin{subfigure}{\linewidth}
\centering
\resizebox{0.9\textwidth}{!}{%
\begin{circuitikz}
\tikzstyle{every node}=[font=\LARGE]
\draw  (1,1) rectangle (2,2);
\draw  (2,1) rectangle (3,2);
\draw  (3,1) rectangle (4,2);
\draw  (4,1) rectangle (5,2);
\draw  (5,1) rectangle (6,2);
\draw  (6,1) rectangle (7,2);
\draw  (7,1) rectangle (8,2);
\draw  (8,1) rectangle (9,2);
\node [font=\LARGE] at (1.5,1.5) {1};
\node [font=\LARGE] at (2.5,1.5) {2};
\node [font=\LARGE] at (3.5,1.5) {3};
\node [font=\LARGE] at (4.5,1.5) {4};
\node [font=\LARGE] at (5.5,1.5) {5};
\node [font=\LARGE] at (6.5,1.5) {6};
\node [font=\LARGE] at (7.5,1.5) {7};
\node [font=\LARGE] at (8.5,1.5) {8};

\draw [->, >=Stealth] (3.5,3) -- (3.5,2);
\draw [->, >=Stealth] (9.5,1.5) -- (10.5,1.5);

\draw  (11,1) rectangle (14,2);
\draw  (14,1) rectangle (19,2);
\node [font=\LARGE] at (12.5,1.5) {1};
\node [font=\LARGE] at (16.5,1.5) {2};
\end{circuitikz}
}%
\caption{The discrete value 3 was selected as the random cut point, so the bins to the left of and including bin 3 were merged into one larger bin and the remaining bins into a second bin.}
\label{fig:rkl_divergence_transformation_basic}
\end{subfigure}

\medskip
\begin{subfigure}{\linewidth}
\centering
\resizebox{0.9\textwidth}{!}{%
\begin{circuitikz}
\tikzstyle{every node}=[font=\LARGE]
\draw  (1,1) rectangle (2,2);
\draw  (2,1) rectangle (3,2);
\draw  (3,1) rectangle (4,2);
\draw  (4,1) rectangle (5,2);
\draw  (5,1) rectangle (6,2);
\draw  (6,1) rectangle (7,2);
\draw  (7,1) rectangle (8,2);
\draw  (8,1) rectangle (9,2);
\node [font=\LARGE] at (1.5,1.5) {1};
\node [font=\LARGE] at (2.5,1.5) {2};
\node [font=\LARGE] at (3.5,1.5) {3};
\node [font=\LARGE] at (4.5,1.5) {4};
\node [font=\LARGE] at (5.5,1.5) {5};
\node [font=\LARGE] at (6.5,1.5) {6};
\node [font=\LARGE] at (7.5,1.5) {7};
\node [font=\LARGE] at (8.5,1.5) {8};

\draw [->, >=Stealth] (3.5,3) -- (3.5,2);
\draw [->, >=Stealth] (9.5,1.5) -- (10.5,1.5);

\draw  (11,1) rectangle (14,2);
\draw  (14,1) rectangle (17,2);
\draw  (17,1) rectangle (18,2);
\draw  (18,1) rectangle (19,2);
\node [font=\LARGE] at (12.5,1.5) {1};
\node [font=\LARGE] at (15.5,1.5) {2};
\node [font=\LARGE] at (17.5,1.5) {3};
\node [font=\LARGE] at (18.5,1.5) {4};
\end{circuitikz}
}%
\caption{Transformation used when evaluating randomised KL divergence in the training pipeline. Two fixed upper-tail boundary points are retained after the randomised aggregation, leaving four bins in total.}
\label{fig:rkl_divergence_transformation_two_dedicated_upper_bins}
\end{subfigure}
\caption{Randomised KL divergence bin-aggregation variants.}
\label{fig:rkl_divergence_transformations}
\end{figure}

We refer to the resulting value as the \textit{randomised KL divergence} (rKL). The basic random split-and-merge step is illustrated in Figure~\ref{fig:rkl_divergence_transformations}\subref{fig:rkl_divergence_transformation_basic}.

For heavy-tailed distributions, we further retain a small number of dedicated upper-tail bins while randomising the coarser interior partition. This preserves resolution where tail behaviour is most important while still reducing variance in the estimation.

For our dataset, we found that keeping the largest and second-largest values as dedicated tail bins produced the most stable training behaviour, corresponding to Figure~\ref{fig:rkl_divergence_transformations}\subref{fig:rkl_divergence_transformation_two_dedicated_upper_bins}. For the 4-qubit-per-variable model, this reduces the effective number of bins per variable from 16 to 4 during loss evaluation.

\paragraph{Matrix formulation.}
Randomised KL divergence can be written as a sequence of matrix products, which keeps every operation differentiable and allows the whole loss to be compiled and vectorised. Let $n$ be the number of qubits per variable and $d$ the number of variables. For a cut point $c \in \{0, \ldots, 2^n - 3\}$, define the aggregation matrix
\[
M^{(c)} = \begin{pmatrix} \mathbf{1}_{[0,\,c)}^\top \\ \mathbf{1}_{[c,\,2^n-3]}^\top \\ e_{2^n-2}^\top \\ e_{2^n-1}^\top \end{pmatrix} \in \{0,1\}^{4 \times 2^n},
\]
whose rows sum, respectively, the bins below the cut, the bins from the cut up to the third-largest value, and the two dedicated tail bins holding the second-largest and the largest values. Here $\mathbf{1}_{S}$ is the indicator vector of the set of bins $S$ and $e_{j}$ is the $j$-th standard basis vector. The aggregation over all variables is the Kronecker product
\[
C_k = \bigotimes_{v=1}^{d} M^{(c_{k,v})} \in \{0,1\}^{4^d \times 2^{nd}},
\]
which maps a distribution over the $2^{nd}$ basis states to one over $4^d$ aggregate cells. Drawing $K$ such configurations, with each cut point $c_{k,v}$ sampled uniformly and independently, the loss is
\[
\mathrm{rKL}(q_\theta, \phi) = \frac{1}{K} \sum_{k=1}^{K} D^\epsilon_{KL}\left( C_k q_\theta,\; C_k \phi \right),
\]
where $q_\theta$ is the model distribution (Section~\ref{sec:amplitude-objective}) and $\phi$ is the empirical distribution. We use $K = 10$. The $K$ configurations are sampled once at the start of training and then held fixed, so each $C_k$ is a constant linear map and the gradient of the loss follows directly from the chain rule through $q_\theta$.

Each $C_k$ is stored as a dense matrix, so memory grows as $K \cdot 4^d \cdot 2^{nd}$. For $d = 4$, $n = 4$ and $K = 10$ this is roughly 700\,MB, which fits comfortably in memory; larger circuits would instead require applying each $M^{(c_{k,v})}$ to its own tensor axis of $q_\theta$.

\begin{figure}
    \includegraphics[width=\linewidth]{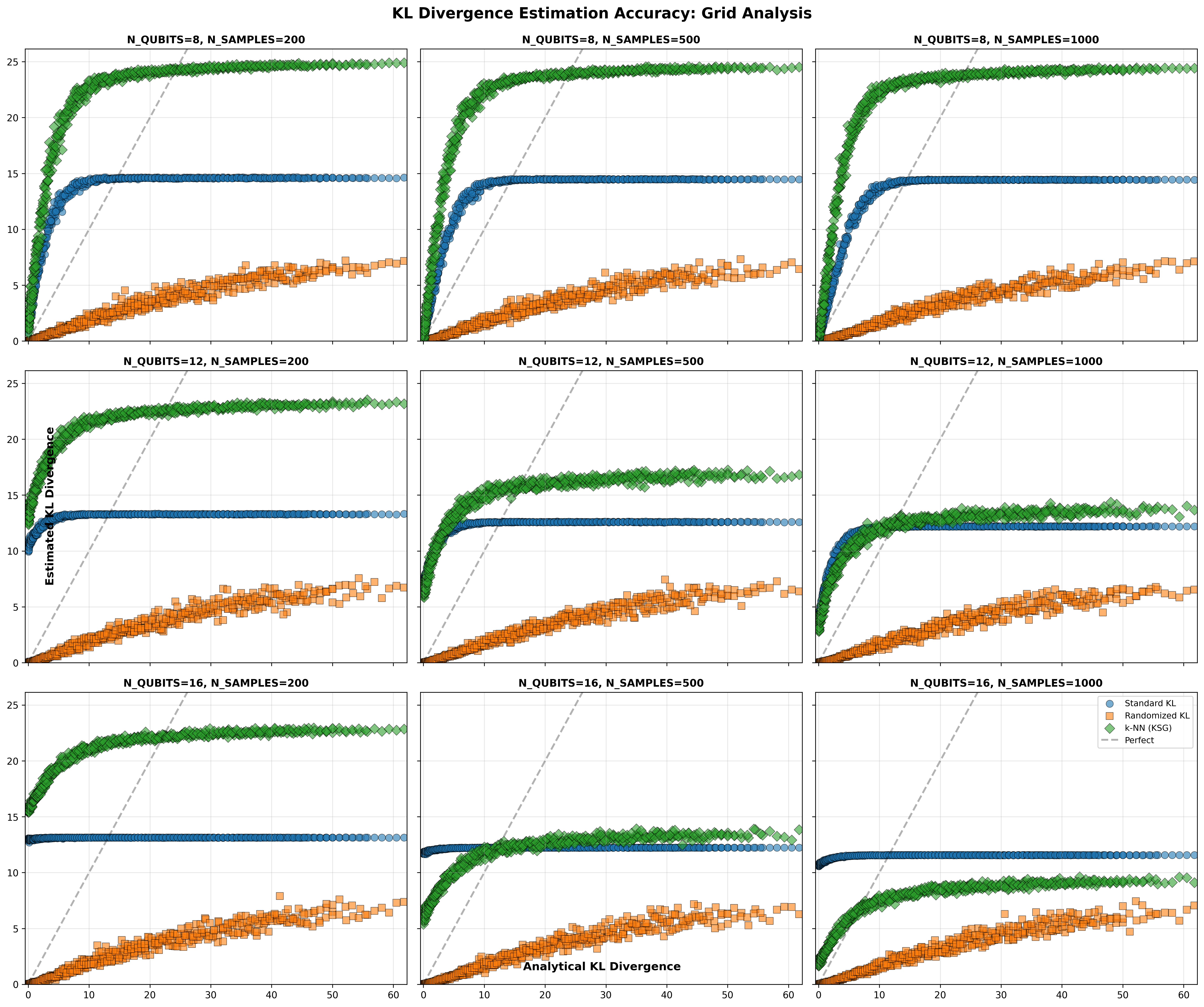}
    \caption{Estimated versus analytical KL divergence for three estimators (standard histogram-based KL, randomised KL, and a $k$-nearest-neighbour estimator) across a grid of discretisation resolutions ($2^8$, $2^{12}$, $2^{16}$ bins, increasing top to bottom) and sample sizes (200, 500, 1000, increasing left to right). For each panel, $n$ samples are drawn from each of $\mathcal{N}(0, 1)$ and $\mathcal{N}(\mu_2, \sigma_2^2)$ and binned into equal-width bins over the common range $[-4.4, 14.4]$; empty bins are set to $10^{-8}$ and the histograms are renormalised to give two discrete probability vectors, from which each estimator is computed. Each point is one configuration $(\mu_2, \sigma_2, \text{repetition})$: its horizontal position is the closed-form KL divergence between the two Gaussians and its vertical position is the estimator's value. The dashed diagonal indicates perfect recovery.}
    \label{fig:rkl_estimates_vs_analytical}
\end{figure}

\begin{figure}
    \includegraphics[width=\linewidth]{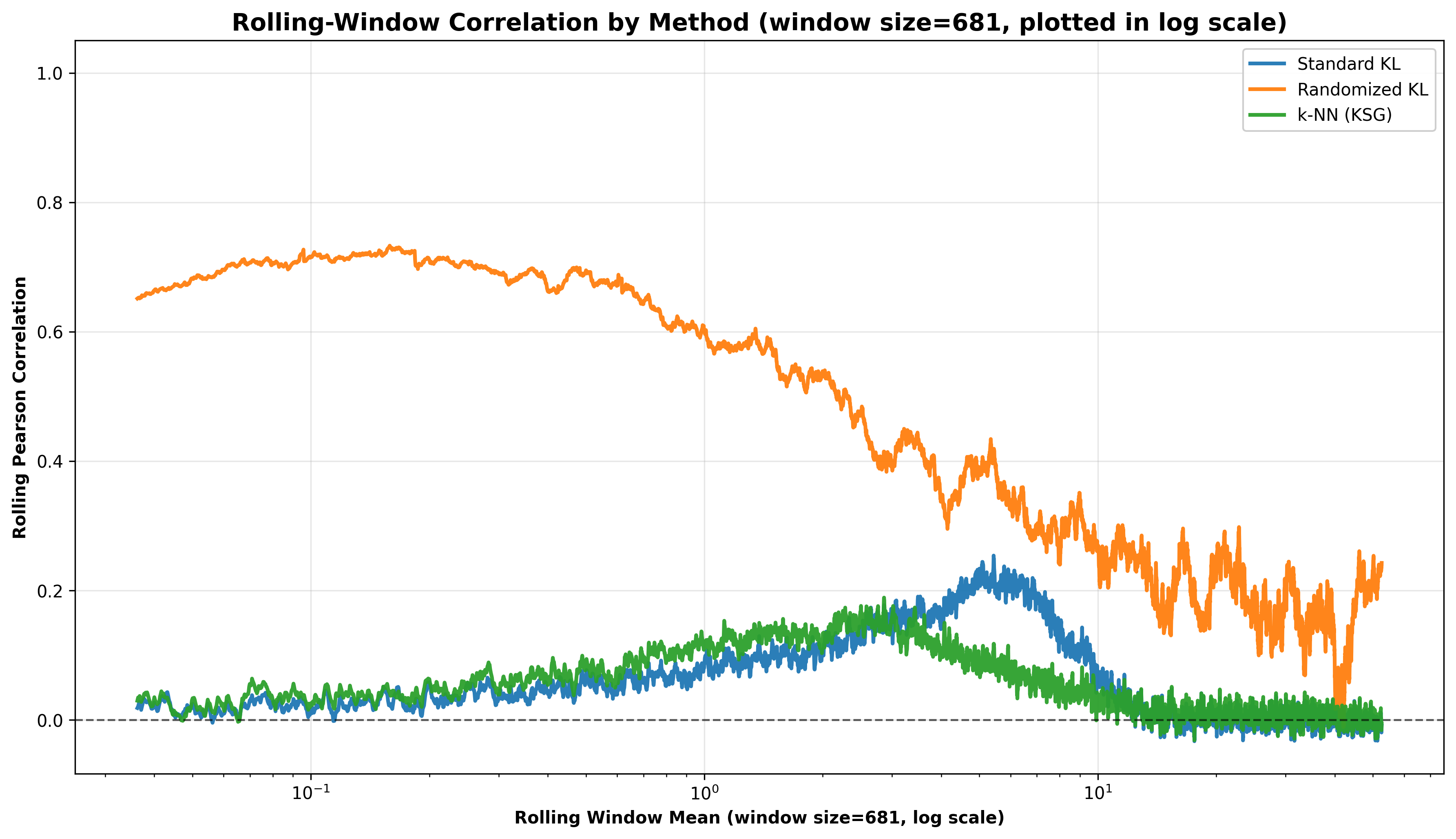}
    \caption{Rolling-window Pearson correlation between each estimator's output and the true analytical KL divergence, computed over all synthetic Gaussian-pair experiments and plotted against the rolling-window mean of the analytical KL (log scale). A correlation near~1 at a given KL magnitude indicates that the estimator reliably ranks model quality in that regime; a value near~0 or negative means the signal is uninformative.}
    \label{fig:rolling_correlation_kl_estimate}
\end{figure}

\subsubsection{Empirical comparison of KL estimation methods}

To motivate the choice of randomised KL divergence as the training objective, we benchmark it against the standard (exact histogram-based) KL divergence and the $k$-nearest-neighbour divergence estimator of \citet{wang2009divergence} across a controlled synthetic setting. Pairs of one-dimensional Gaussians $\mathcal{N}(\mu_1, \sigma_1^2)$ and $\mathcal{N}(\mu_2, \sigma_2^2)$ are drawn for a grid of mean offsets ($\mu_2 \in [0, 10]$) and standard deviations ($\sigma_2 \in \{0.90, 0.95, 1.00, 1.05, 1.10\}$), for which the true KL divergence is available in closed form. Each configuration is repeated three times with fresh samples, and the experiment is repeated across three discretisation resolutions and three sample sizes (200, 500, and 1000 observations). Because the benchmark distribution is one-dimensional, the qubit counts here refer to qubits per single axis; to map these to the main four-variable portfolio setting one should divide by four. Specifically, 8, 12, and 16 qubits in the benchmark correspond to 2, 3, and 4 qubits per variable respectively in the portfolio circuit, so the 16-qubit, 1000-sample condition is the closest analogue to the realistic training scenario, which has 939 observations.

Figure~\ref{fig:rkl_estimates_vs_analytical} shows estimated versus analytical KL divergence for each method across all grid cells. Each panel corresponds to a (discretisation, sample-size) combination. Points on the diagonal indicate exact recovery. The standard KL estimator tracks the diagonal closely only when the number of samples is large relative to the number of bins; at finer discretisation or smaller sample sizes---precisely the regime of the main problem---most bins receive no observations, the empirical distribution becomes unreliable, and the estimate degrades severely. The $k$-NN estimator avoids binning but instead operates on a finite sample drawn from the histogram, which introduces additional variance and similarly degrades when samples are scarce. The randomised KL estimator is systematically biased below the diagonal---it underestimates the true divergence because aggregating bins erases within-bin variation---but its estimates remain monotonically ordered with the analytical values across all conditions tested.

Figure~\ref{fig:rolling_correlation_kl_estimate} translates this observation into a training-relevant metric. For each estimator, we sort all evaluation instances by analytical KL divergence and compute the Pearson correlation between the estimator's output and the analytical value within a rolling window centred at each point. The result is plotted on a log scale along the horizontal axis to resolve the behaviour across the several orders of magnitude spanned by the analytical values. A correlation near one at a given KL magnitude means that the estimator reliably ranks model quality in that regime; a correlation near zero or negative means the signal is uninformative or misleading.

Only the randomised KL estimator maintains high rolling correlations consistently across the full range of analytical divergences. Both the standard KL and the $k$-NN estimator lose correlation in the sparse-data regime: standard KL breaks down because most discretisation bins are empty when observations are few relative to the support size, making the empirical probability vector a poor proxy for the true distribution; the $k$-NN estimator similarly degrades when nearest-neighbour distances become uninformative due to repeated draws from the same histogram bins. For gradient-based training the absolute scale of the loss is irrelevant: what matters is that the estimator correctly identifies which parameter perturbation reduces the divergence. The randomised KL divergence achieves this by collapsing the support to just a handful of aggregate bins, drastically reducing the number of empty cells and restoring a reliable signal even at sample sizes that are one or two orders of magnitude smaller than the support. It therefore provides a robust directional signal for stochastic gradient descent in the sparse-data regime that characterises insurance portfolio copula fitting.

This benchmark has three limitations. First, it is one-dimensional, and the Gaussian pairs differ only in location and scale, which coarse aggregate bins detect easily; it does not test the estimator's sensitivity to differences in multivariate dependence, which is what the loss must capture when fitting a copula. Second, the benchmark uses 100 random cuts without a dedicated tail bin, whereas training uses 10 cuts with dedicated tail bins. Third, because randomised KL divergence compares distributions only through their aggregate cells, two distributions that agree on every aggregate cell of the sampled cuts have zero randomised KL divergence even if they differ at full resolution. The loss is therefore a coarse proxy for the full KL divergence: it trades resolution for a stable training signal.

\subsection{Hierarchical learning}\label{sec:hierarchical-learning}

In \cite{zhu2023copula} the authors rely on SPSA~\cite{original_spsa_paper} for training, which estimates the gradient from two circuit evaluations per step, regardless of the number of parameters. Exact parameter-shift gradients~\citep{mitarai2018quantum,schuld2019evaluating} are also available on hardware, but they require two circuit evaluations per parameter per step, which is costly at the number of parameters and shots involved here. In either case, gradients estimated from sampled circuits are noisy and expensive to obtain, which substantially limits training efficiency. 

We instead adopt a hybrid strategy based on the hierarchical (coarse-to-fine) training of \citet{gharibyan2023hierarchical}, in which qubits are added to a QCBM to refine the discretisation and the parameters of the smaller circuit serve as a warm start for the larger one. Gharibyan et al.\ train their circuits in simulation and then execute the trained circuits on IBM superconducting hardware; our contribution is to apply the approach to learning copulas for risk aggregation, and to fine-tune only the newly added parameters with SPSA under a realistic trapped-ion noise model. A smaller model, using the same ansatz as in \cite{zhu2023copula}, is first trained on a classical state-vector simulator, where gradients are efficiently computed and the model can be better trained. The learned parameters are then embedded into a larger model, after which SPSA is used only for fine-tuning on the quantum hardware. A minor modification of the ansatz is required to make this embedding precise.

\begin{figure}
    \centering
    \begin{subfigure}{\linewidth}
        \centering
        \begin{quantikz}
            \lstick{$\ket{q^{(1)}_0}$} & \gate{H} & \ctrl{3} & \qw      & \qw      & \qw \\
            \lstick{$\ket{q^{(1)}_1}$} & \gate{H} & \qw      & \ctrl{3} & \qw      & \qw \\
            \lstick{$\ket{q^{(1)}_2}$} & \gate{H} & \qw      & \qw      & \ctrl{3} & \qw \\
            \lstick{$\ket{q^{(j)}_0}$} & \qw      & \targ{}  & \qw      & \qw      & \qw \\
            \lstick{$\ket{q^{(j)}_1}$} & \qw      & \qw      & \targ{}  & \qw      & \qw \\
            \lstick{$\ket{q^{(j)}_2}$} & \qw      & \qw      & \qw      & \targ{}  & \qw
        \end{quantikz}
        \caption{Fixed state preparation, shown for three qubits per variable: Hadamard gates on the register of the first variable, with CNOTs copying each qubit to the corresponding qubit of variable $j$, for each $j = 2, 3, 4$. With four qubits per variable, $q_3$ is included in the same way.}
        \label{fig:state_prep}
    \end{subfigure}

    \vspace{0.8\baselineskip}

    \begin{subfigure}{\linewidth}
        \centering
        \begin{quantikz}
            \lstick{$\ket{q_0}$} & \gate{Z_0} & \gate{X_0} & \gate{RZZ_0}\vqw{1} & \qw & \gate{RZZ_2}\vqw{2} & \qw \\
            \lstick{$\ket{q_1}$} & \gate{Z_1} & \gate{X_1} & \gate{RZZ_0} & \gate{RZZ_1}\vqw{1} & \qw      & \qw \\
            \lstick{$\ket{q_2}$} & \gate{Z_2} & \gate{X_2} & \qw      & \gate{RZZ_1} & \gate{RZZ_2} & \qw
        \end{quantikz}
        \caption{One parameterised layer of the smaller model on a single variable's three-qubit register, featuring single-qubit $Z$ and $X$ rotations and two-qubit $RZZ$ entangling gates.}
        \label{fig:smaller_model}
    \end{subfigure}

    \vspace{0.8\baselineskip}

    \begin{subfigure}{\linewidth}
        \centering
        \begin{quantikz}
            \lstick{$\ket{q_0}$} & \gate{Z_0}\gategroup[3,steps=5,style={dashed,rounded corners,fill=blue!10, inner xsep=2pt},background,label style={label position=below,anchor=north,yshift=-0.2cm}]{} & \gate{X_0} & \gate{RZZ_0}\vqw{1} & \qw                 & \gate{RZZ_2}\vqw{2} & \qw                 & \gate{RZZ_4}\vqw{3} & \qw \\
            \lstick{$\ket{q_1}$} & \gate{Z_1} & \gate{X_1} & \gate{RZZ_0}        & \gate{RZZ_1}\vqw{1} & \qw                 & \qw                 & \qw          & \qw \\
            \lstick{$\ket{q_2}$} & \gate{Z_2} & \gate{X_2} & \qw                 & \gate{RZZ_1}        & \gate{RZZ_2}        & \gate{RZZ_3}\vqw{1} & \qw          & \qw \\
            \lstick{$\ket{q_3}$} & \gate{Z_3} & \gate{X_3} & \qw                 & \qw                 & \qw                 & \gate{RZZ_3}        & \gate{RZZ_4} & \qw
        \end{quantikz}
        \caption{One parameterised layer of the larger model on a single variable's four-qubit register, which contains the smaller layer (shaded) on the first three qubits.}
        \label{fig:embedded_model}
    \end{subfigure}
    \caption{Hierarchical-learning ansatzes used in this work: fixed state preparation (top), followed by two repetitions of the parameterised layer applied to each variable's register, for the smaller pretrained circuit (middle) and the larger embedded circuit (bottom).}
    \label{fig:hierarchical_learning_ansatzes}
\end{figure}
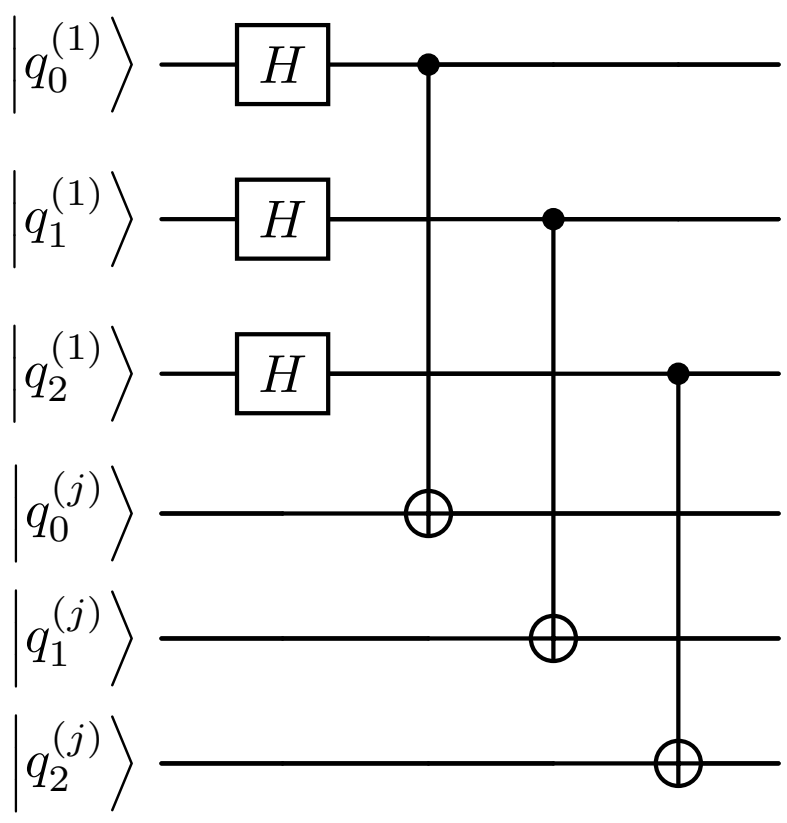
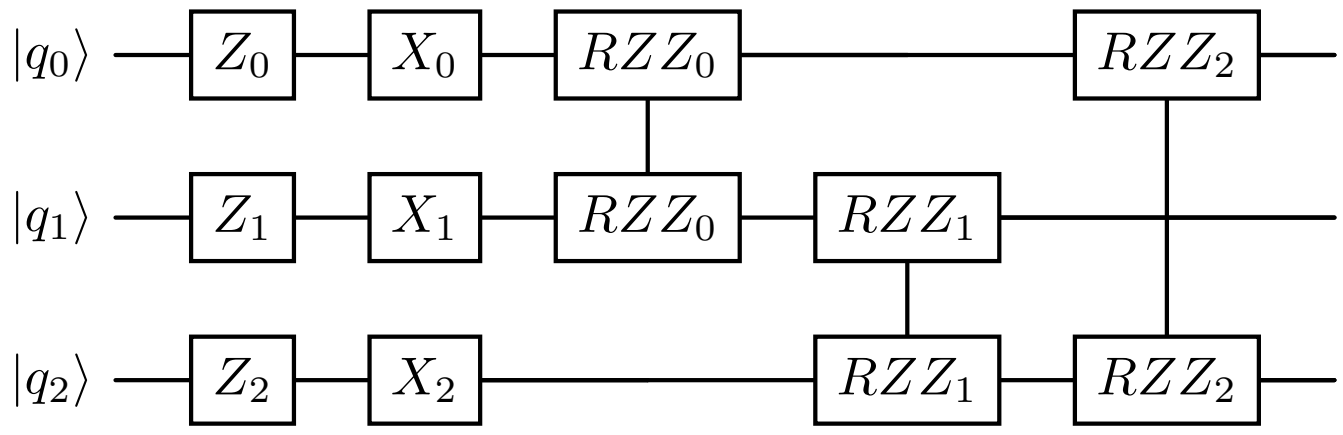
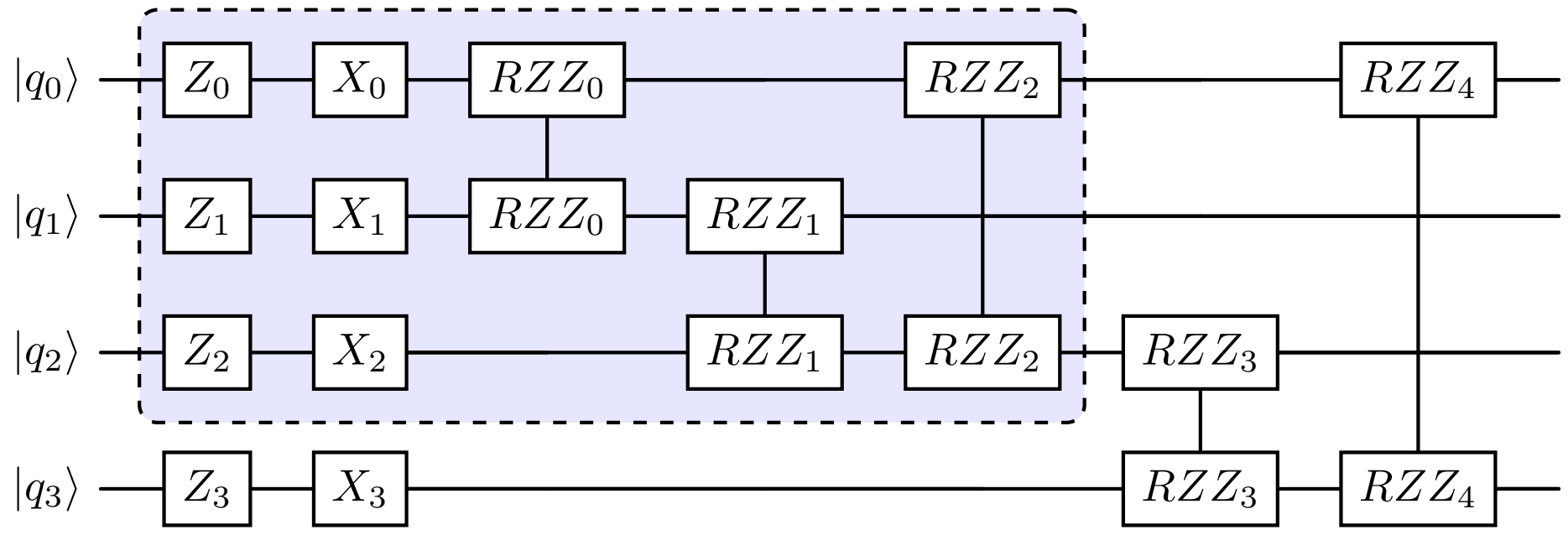

When a new qubit is added to the circuit, the smaller circuit appears as a subcircuit of the larger one. The retained $RZZ_2$ gate (Figure~\ref{fig:hierarchical_learning_ansatzes}\subref{fig:embedded_model}) makes the enlarged ansatz differ slightly from the original design, but it is preserved so that the transferred model reproduces the smaller model at initialisation.
Before the parameterised layers, a fixed state-preparation step applies a Hadamard gate to each qubit of the first variable's register, followed by a CNOT from each of these qubits to the corresponding qubit of every other variable's register (Figure~\ref{fig:hierarchical_learning_ansatzes}\subref{fig:state_prep}). These CNOTs are the only gates that couple different variables; every parameterised gate acts within a single variable's register. The parameterised layer is applied twice.
In the larger model, the original encoding is preserved and a further uniform split is added through the new qubit. Equivalently, for each variable, a binary string $b_1\ldots b_n$ is mapped to either $b_1\ldots b_n0$ or $b_1\ldots b_n1$ with equal probability. Because the state-preparation CNOTs copy the new qubit of the first variable to the new qubit of every other variable, these new bits are perfectly correlated across variables at initialisation: all are 0 or all are 1, each with probability one half. Subsequent training of the new parameters then adjusts the joint distribution of the refinement bits.

The new parameters $Z_3$, $X_3$, $RZZ_3$, and $RZZ_4$ are initialised to zero, and only these newly introduced parameters are optimised with SPSA. This reduces the dimensionality of the optimisation problem, which makes each SPSA gradient estimate less noisy, while keeping the transferred parameters fixed.

\section{Training Pipeline}
\label{sec:training-pipeline}

Our training pipeline has two stages. First, we train a smaller model with three qubits per variable on a local state-vector simulator using gradient-based optimisation. Second, we embed the learned parameters into a larger model with four qubits per variable and fine-tune only the new parameters on quantum hardware using SPSA. 

In the first stage, the empirical data is encoded with the exponential quantile scheme described in Section~\ref{sec:methods}, yielding three qubits per variable and therefore 12 qubits across the four-variable portfolio. We choose a starting point by random search over initial parameter settings, tune the learning rate with Optuna~\cite{optuna}, and then train all parameters with Adam, using exact gradients from the state-vector simulator and stopping early once the loss no longer improves. Throughout this stage, the loss is the randomised KL divergence averaged over 10 random cuts (see Appendix~\ref{app:details} for all settings).

In the second stage, the encoding is expanded to four qubits per variable by adding one uniform bit per variable, thereby splitting each previous bin into two equal sub-bins as described in Section~\ref{sec:hierarchical-learning}. The learned weights are embedded into the larger circuit, and the new parameters are initialised to zero. Only these newly introduced parameters are trained with SPSA on the target hardware, using a learning rate of 0.06, a perturbation size of 0.002, and blocking to reject updates that increase the loss. Each gradient estimate is averaged over five resamplings. The reported parameters are those with the lowest loss among all points evaluated during optimisation, including the perturbed points used for gradient estimation.

Every circuit execution on the H1 emulator or on H1-1 uses 1000 shots. SPSA fine-tuning on the H1 emulator runs for 100 iterations; with five resamplings and blocking, each iteration requires 11 circuit executions (ten for the gradient estimate and one to evaluate the proposed update), giving roughly 1100 circuit executions, or about $1.1 \times 10^{6}$ shots in total. The fine-tuned parameters are then evaluated on H1-1 with a single 1000-shot execution. On the state-vector simulator, the model distribution is computed exactly from the amplitudes, so the simulator stages involve no shots; the distributions shown in Section~\ref{sec:results} are built from 1000 samples drawn from each model.

\subsection{Amplitude-space training objective}
\label{sec:amplitude-objective}

The standard QML approach is to define the loss on the Born probabilities $|\psi_\theta(x)|^2$ of the circuit state $\ket{\psi_\theta}$. We instead define the model distribution on the amplitude magnitudes,
\[
q_\theta(x) := \frac{|\psi_\theta(x)|}{\sum_{y} |\psi_\theta(y)|},
\]
and compute the randomised KL divergence between $q_\theta$ and the empirical distribution $\phi$ over basis states, rather than between the Born probabilities and $\phi$. Compared with the Born probabilities, $q_\theta$ is less sharply peaked, and in practice this gave a more favourable optimisation landscape.

On the state-vector simulator, $q_\theta$ is computed exactly from the amplitudes. On the emulator and on hardware, where only measurement counts are available, it is estimated from the observed frequencies $\hat p$ as $\sqrt{\hat p(x)} / \sum_y \sqrt{\hat p(y)}$. Samples from the trained model, used in all evaluations below, are drawn from $q_\theta$.

\section{Results}
\label{sec:results}

We evaluate the method on two datasets:
\begin{enumerate}
\item The first, denoted ``Stock/Insurance,'' contains weekly Nikkei 225 negative returns (i.e. losses) together with three sources of insurance loss. The loss lines are positively dependent on one another, whereas their dependence on the equity returns is weak, close to independence, including in the upper tail.
\item The second, denoted ``Gumbel,'' is a synthetic Gumbel copula included to test the training procedure on a standard benchmark distribution using the usual marginal-normalisation workflow.
\end{enumerate}

For both datasets, we evaluate the aggregate obtained by summing the four variables with equal weight. In the Gumbel benchmark, all four variables are on the same scale. In the Stock/Insurance dataset, the standardised Nikkei~225 returns are small compared with the insurance losses in the upper tail, so the upper tail of the aggregate depends on how often large losses occur together.

\begin{figure}
  \centering
  \begin{subfigure}{0.45\textwidth}
    \includegraphics[width=\linewidth]{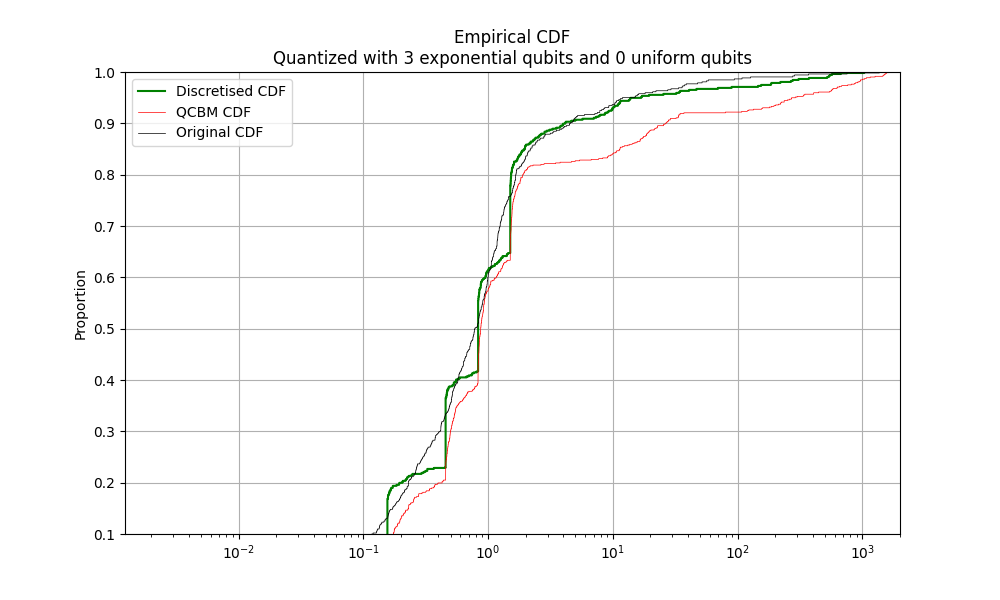}
    \caption{Stock/Insurance aggregate, local simulator, 3 qubits per variable}
    \label{fig:stockloss_local_aggregate_qpv3}
  \end{subfigure}
  \begin{subfigure}{0.45\textwidth}
    \includegraphics[width=\linewidth]{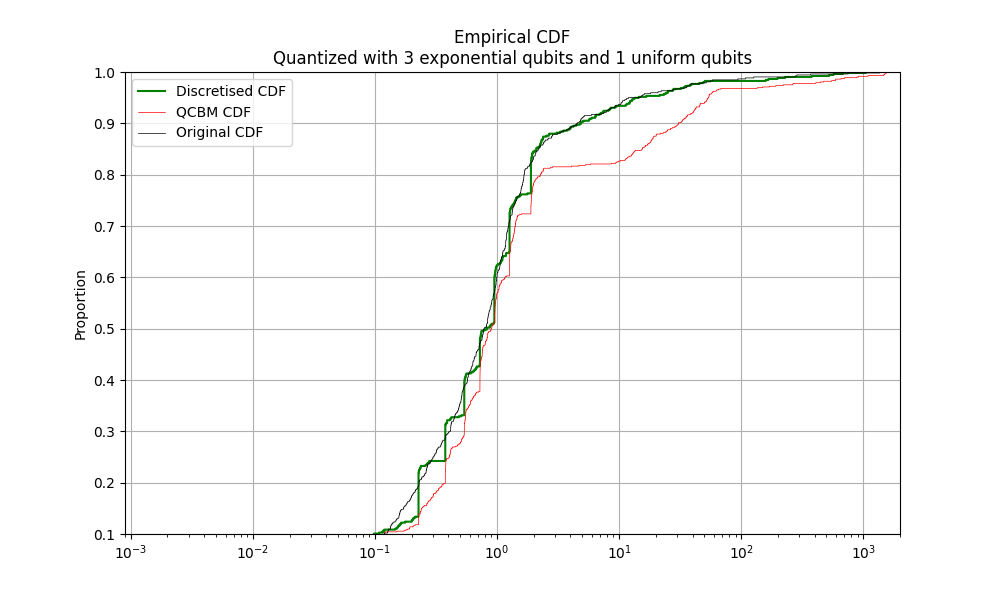}
    \caption{Stock/Insurance aggregate after transfer to 4 qubits per variable}
    \label{fig:stockloss_local_aggregate_qpv4}
  \end{subfigure}
  \begin{subfigure}{0.45\textwidth}
    \includegraphics[width=\linewidth]{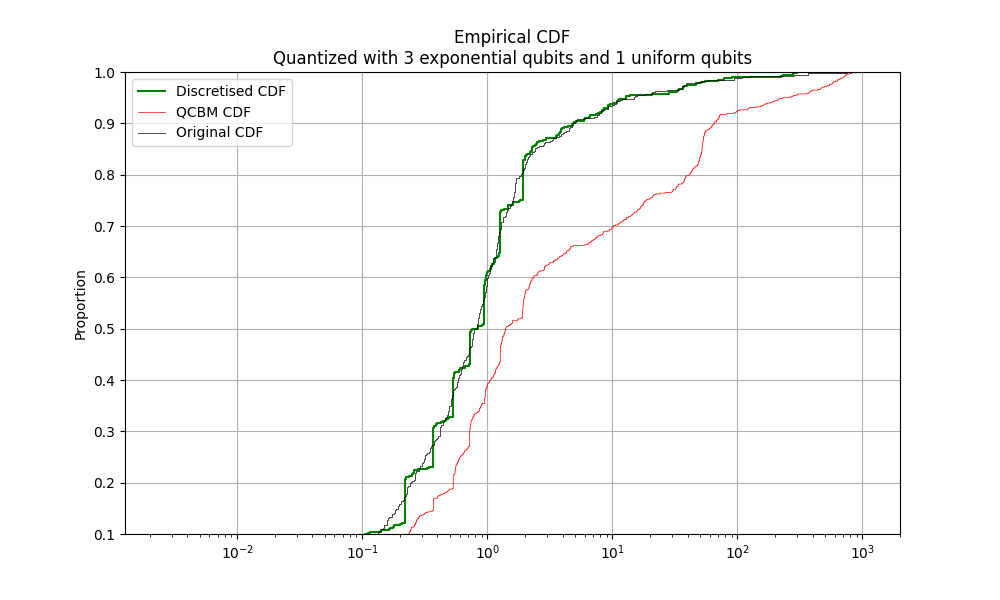}
    \caption{Stock/Insurance aggregate, 4 qubits per variable after SPSA on the noisy H1 emulator}
    \label{fig:stockloss_h1_emulator_aggregate}
  \end{subfigure}
  \hfill
  \begin{subfigure}{0.45\textwidth}
    \includegraphics[width=\linewidth]{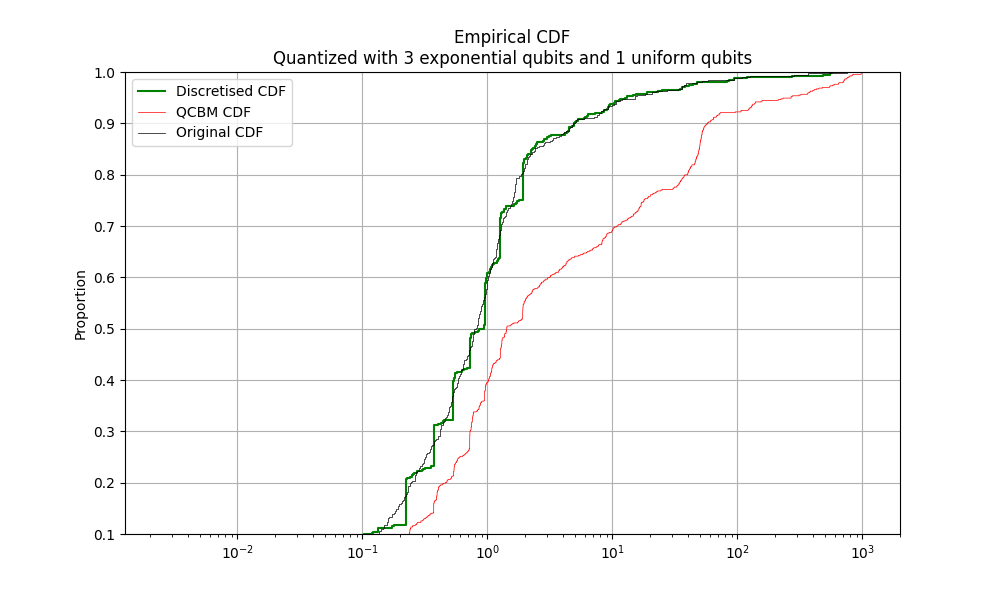}
    \caption{Stock/Insurance aggregate, 4 qubits per variable, emulator-trained SPSA parameters evaluated on Quantinuum H1-1}
    \label{fig:stockloss_h1_1_aggregate}
  \end{subfigure}
  \caption{Cumulative distribution functions for total weekly losses (i.e. the sum over the dimensions) of the Stock/Insurance dataset at the stages of the training pipeline: simulator pretraining at 3 qubits per variable, transfer to 4 qubits per variable, SPSA fine-tuning of the 4-qubit-per-variable model on the noisy H1 emulator, and evaluation of the same fine-tuned parameters on H1-1.}
  \label{fig:stockloss_results}
\end{figure}

Figure~\ref{fig:stockloss_results} summarises the full training pipeline on the Stock/Insurance dataset. The 3-qubit-per-variable model trained on the local state-vector simulator already reproduces the aggregate portfolio distribution reasonably well. After transfer to the 4-qubit-per-variable circuit, the fit remains close to the discretised target, indicating that the transfer step preserves the learned distribution rather than disrupting it.

Hardware experiments were performed on Quantinuum H1-1, a 20-qubit trapped-ion quantum computer based on the quantum charge-coupled device (QCCD) architecture~\cite{Pino_2021}. H1-1 provides all-to-all qubit connectivity, so no qubit routing is needed to entangle the variable registers. Figure~\ref{fig:stockloss_h1_1_aggregate} shows the result on H1-1 with the SPSA-fine-tuned parameters. Evaluated on the noiseless simulator, SPSA fine-tuning reduces the randomised KL divergence of this model from 1.28 at transfer to 1.23, about 3.5\%. With the same parameters, the H1 emulator and H1-1 reach 1.97 and 2.06 respectively. Most of this gap comes from the finite number of shots rather than from hardware noise. The hardware distribution is estimated from 1000 shots spread over $2^{16}=65{,}536$ basis states, so most states are never observed. Sampling 1000 shots from the noiseless model and applying the same square-root post-processing gives a randomised KL divergence of $1.90 \pm 0.06$ (mean and standard deviation over 200 repetitions), which accounts for about 80\% of the gap between the noiseless model and H1-1; with 10{,}000 shots it falls to 1.38. H1-1 therefore adds only about 0.16 to the divergence obtained by ideal sampling with the same number of shots. Resource constraints prevented us from running more shots on hardware, and we expect the hardware results to improve with more shots. Because the number of basis states grows exponentially with the number of variables, however, so does the number of shots needed to estimate the distribution, which is a potential limitation for larger portfolios.

\begin{figure}
  \centering
  \begin{subfigure}{0.45\textwidth}
    \includegraphics[width=\linewidth]{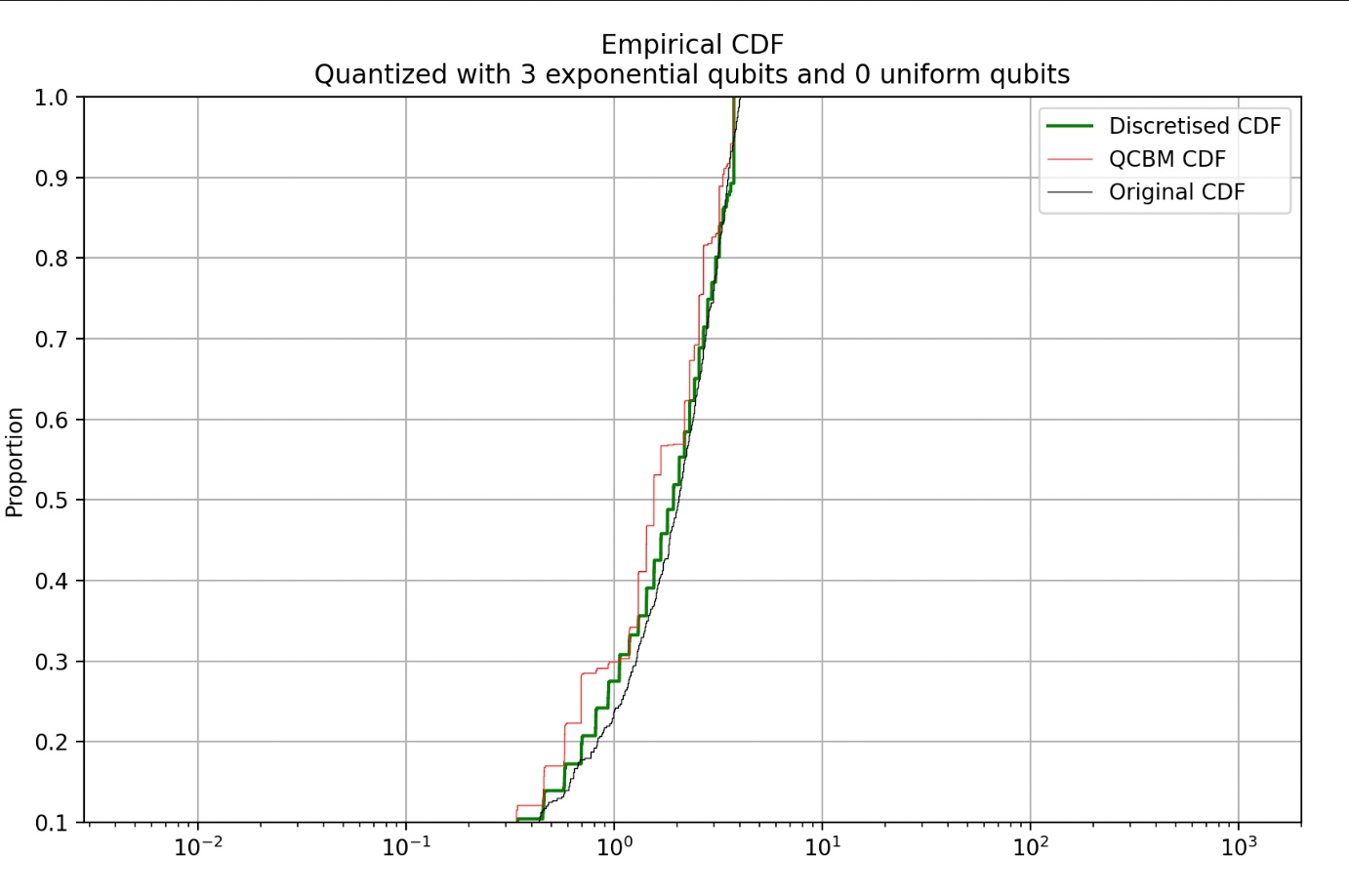}
    \caption{Gumbel aggregate, local simulator, 3 qubits per variable}
    \label{fig:gumbel_local_aggregate_qpv3}
  \end{subfigure}
  \begin{subfigure}{0.45\textwidth}
    \includegraphics[width=\linewidth]{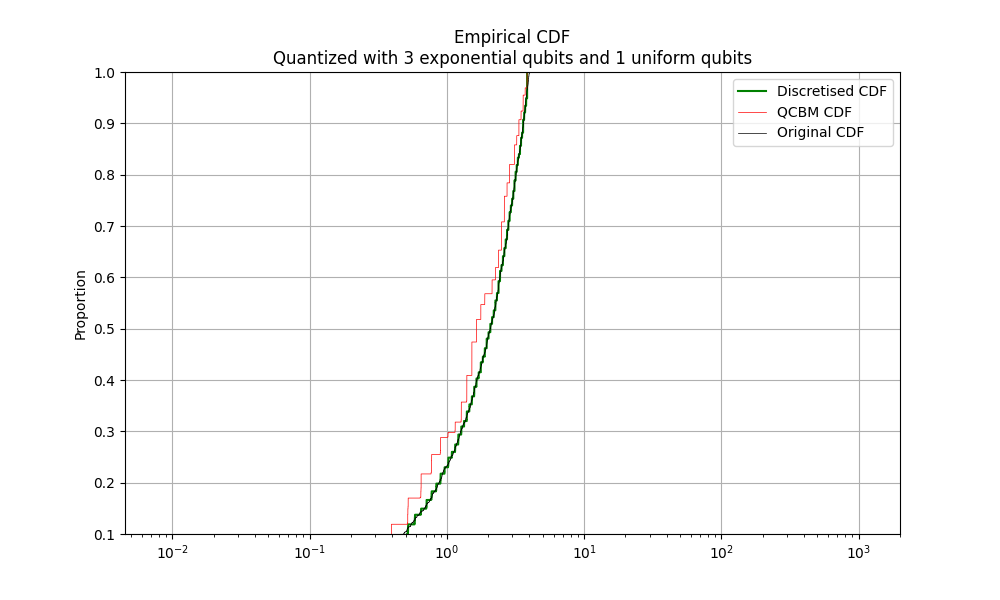}
    \caption{Gumbel aggregate after transfer to 4 qubits per variable}
    \label{fig:gumbel_local_aggregate_qpv4}
  \end{subfigure}
\hfill
  \begin{subfigure}{0.45\textwidth}
    \includegraphics[width=\linewidth]{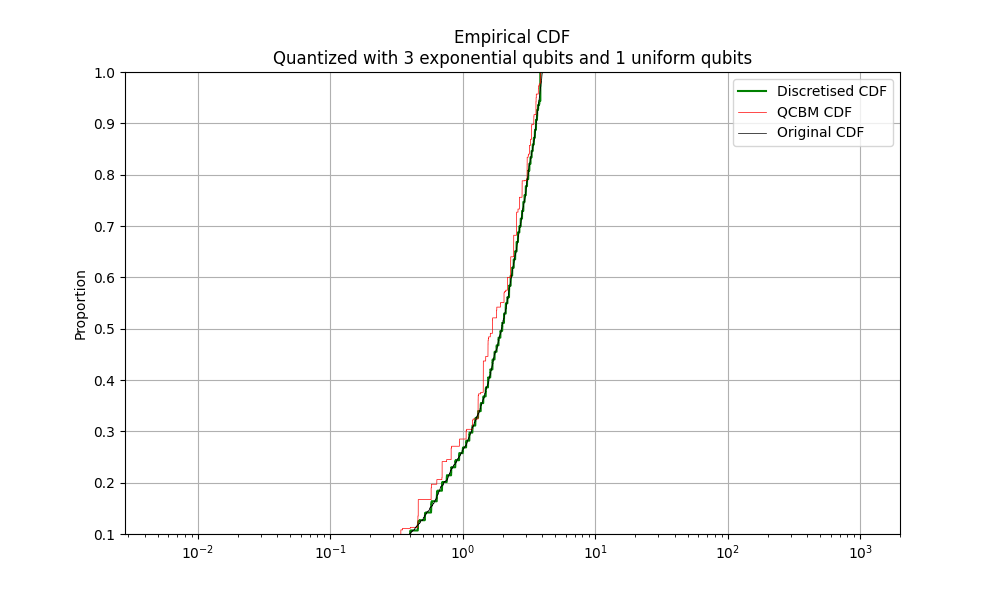}
    \caption{Gumbel aggregate, 4 qubits per variable after SPSA on the noisy H1 emulator}
    \label{fig:gumbel_h1_emulator_aggregate}
  \end{subfigure}
  \hfill
  \begin{subfigure}{0.45\textwidth}
    \includegraphics[width=\linewidth]{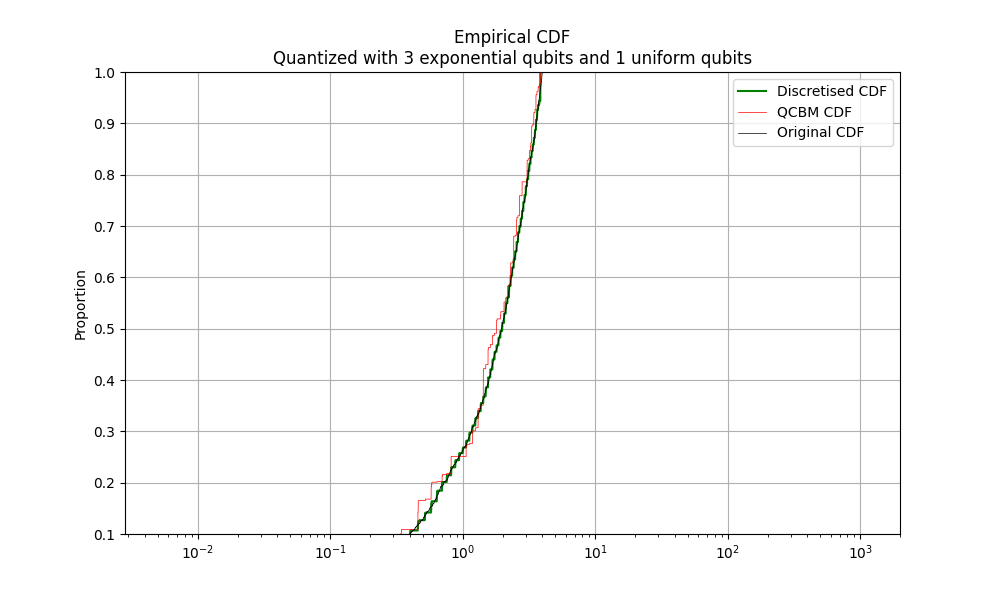}
    \caption{Gumbel aggregate, 4 qubits per variable, SPSA parameters evaluated on Quantinuum H1-1}
    \label{fig:gumbel_h1_1_aggregate}
  \end{subfigure}
  \caption{Cumulative distribution functions for the sum over the dimensions of the Gumbel benchmark across the same training pipeline used for the Stock/Insurance dataset.}
  \label{fig:gumbel_results}
\end{figure}

Figure~\ref{fig:gumbel_results} reports the same pipeline on the Gumbel benchmark. The qualitative behaviour is similar: pretraining and hierarchical learning work well in simulation, and the fit degrades on the emulator and on H1-1.

\subsection{Effectiveness of Hierarchical Learning}

To measure the value of hierarchical learning, we compare several initialisation and optimisation strategies for the 4-qubit-per-variable model. In our setting, hierarchical learning means initialising the larger circuit from the trained smaller model and then optimising only the newly introduced parameters.

The primary motivation for hierarchical learning in our setting is to identify a favourable starting point in the optimisation landscape, which is known to contain barren plateaus~\cite{cite_barren_plateus_are_everywhere} and local minima whose number grows exponentially with the number of qubits~\cite{cite_exponentially_many_local_minima}. By warm-starting with a pre-trained smaller model, we significantly improve convergence compared to random initialisation. Related strategies that grow the circuit during training~\citep{skolik2021layerwise} or choose initial parameters so that new blocks act as the identity~\citep{grant2019initialization} have likewise been proposed to mitigate barren plateaus.

Our transfer strategy also freezes the inherited parameters and updates only the new ones. This preserves the structure already learned at lower resolution while allowing the larger model to refine the discretisation.

\textbf{Empirical Results.} We compared four initialisation and optimisation strategies for our 4-qubit-per-variable model on the Stock/Insurance dataset, using the state-vector simulator. In the random-initialisation arm, the circuit uses the original ansatz of \citet{zhu2023copula} with four qubits per variable, which has 96 parameters (it lacks the retained $RZZ_2$ gate of the embedded ansatz, one per variable in each layer); all of them start from the best of 1000 uniformly random parameter draws, with no gradient training, and SPSA then updates all of them. In the hierarchical arm, the 72 parameters of the pretrained 3-qubit-per-variable model are transferred and frozen, and SPSA updates only the 32 new parameters. Both arms use 300 SPSA iterations (3303 loss evaluations) with identical SPSA settings, and each arm was run once. To make the values comparable, all four models are scored with the training loss, the randomised KL divergence, against the same target: the hierarchical encoding with three exponential bits and one uniform bit per variable.

\begin{itemize}
    \item \textbf{Random initialisation}: randomised KL divergence of 2.41.
    \item \textbf{Random initialisation with SPSA}: SPSA reduced this to 1.42.
    \item \textbf{Hierarchical learning}: the transferred model, before SPSA, has a randomised KL divergence of 0.75.
    \item \textbf{Hierarchical learning with SPSA}: our combined approach achieved the best performance, with a randomised KL divergence of 0.56.
\end{itemize}

\begin{figure}
  \centering
  \begin{subfigure}{0.45\textwidth}
    \includegraphics[width=\linewidth]{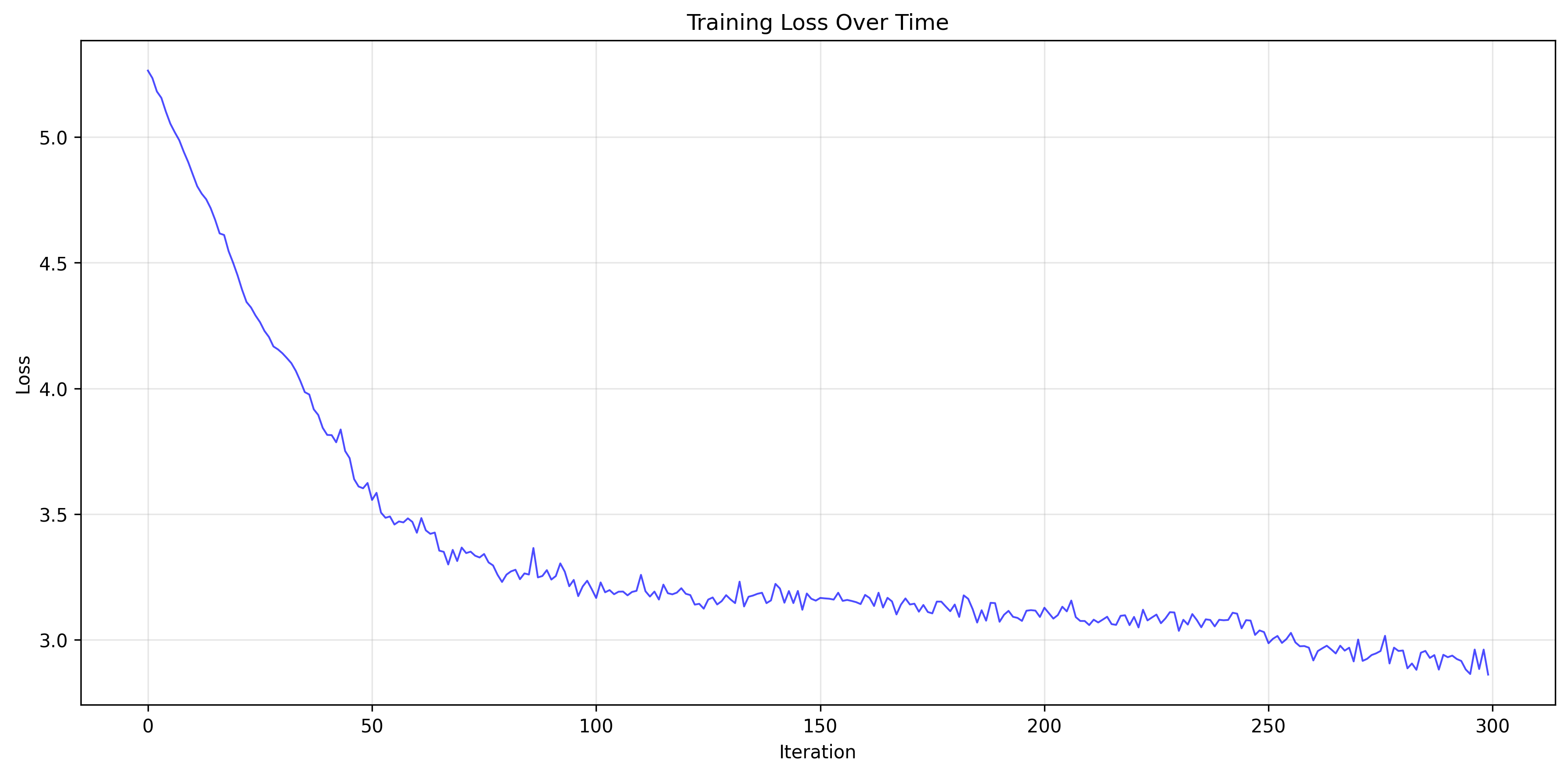}
    \caption{Random initialisation with SPSA}
    \label{fig:no_hierarchical_learning_losses}
  \end{subfigure}
  \begin{subfigure}{0.45\textwidth}
    \includegraphics[width=\linewidth]{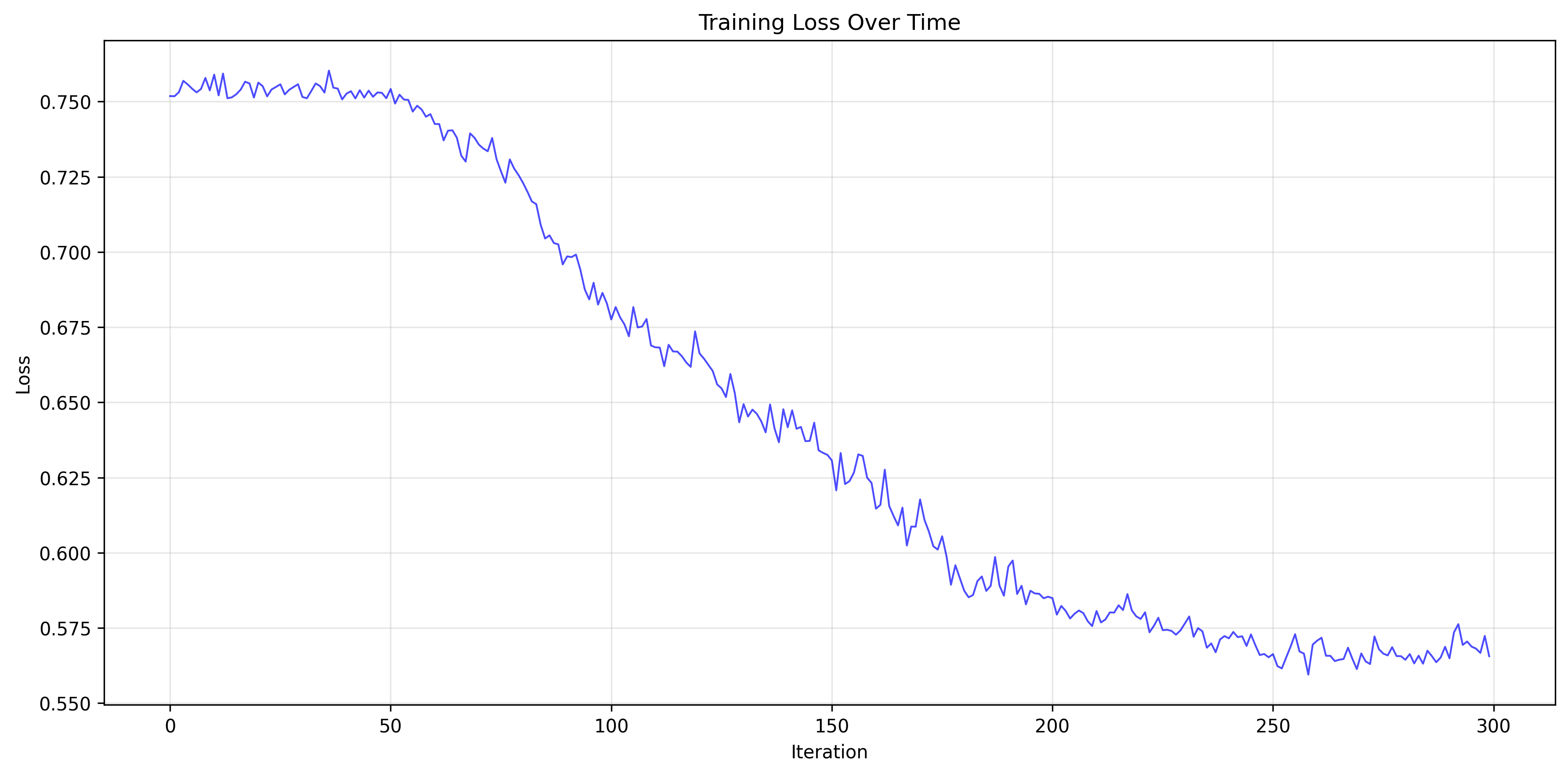}
    \caption{Hierarchical learning with SPSA}
    \label{fig:hierarchical_learning_losses}
  \end{subfigure}
  \caption{SPSA training loss of the 4-qubit-per-variable model, starting from random initialisation (left) or from the transferred 3-qubit-per-variable model (right). The transferred model converges noticeably faster. Each panel shows the randomised KL divergence against its own training target, which has four exponential bits per variable on the left and three exponential bits plus one uniform bit on the right, so the two vertical scales are not directly comparable. Scored against a common target, the transferred model reaches 0.56, against 1.42 for random initialisation.}
  \label{fig:effect_of_hierarchical_learning}
\end{figure}

These results show that hierarchical learning materially improves optimisation. The transferred model alone already outperforms random initialisation followed by SPSA, and relative to that baseline the full hierarchical learning pipeline reduces the final randomised KL divergence by a factor of about 2.5 (1.42 versus 0.56). The comparison is not matched in compute: the hierarchical arm additionally benefits from the gradient-based pretraining of the smaller model. Figure~\ref{fig:effect_of_hierarchical_learning} shows that hierarchical learning also produces faster and more stable convergence.

\begin{figure}
  \centering
  \begin{subfigure}{0.45\textwidth}
    \includegraphics[width=\linewidth]{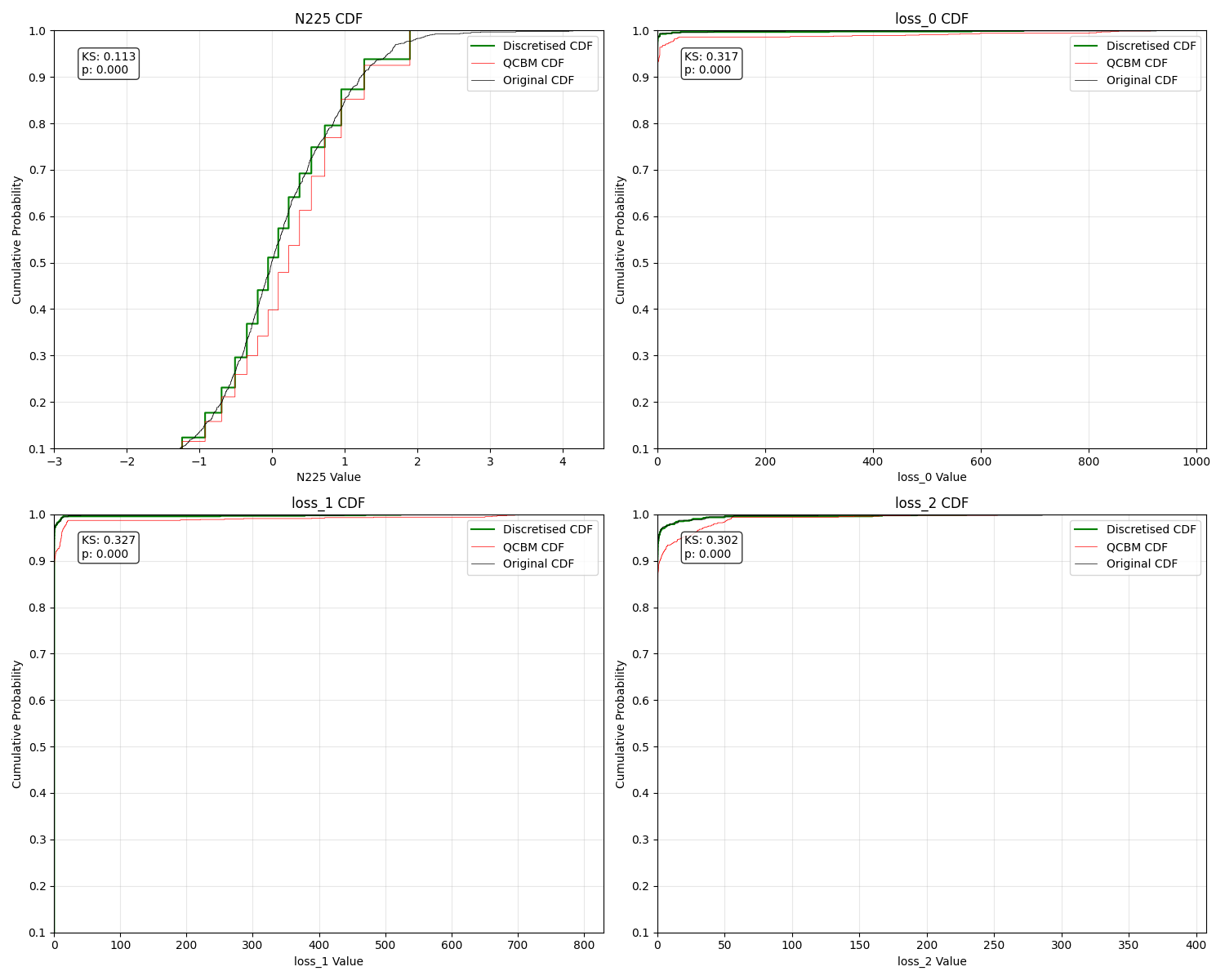}
    \caption{Marginals after hierarchical learning}
    \label{fig:hierarchical_learning_marginals}
  \end{subfigure}
  \begin{subfigure}{0.45\textwidth}
    \includegraphics[width=\linewidth]{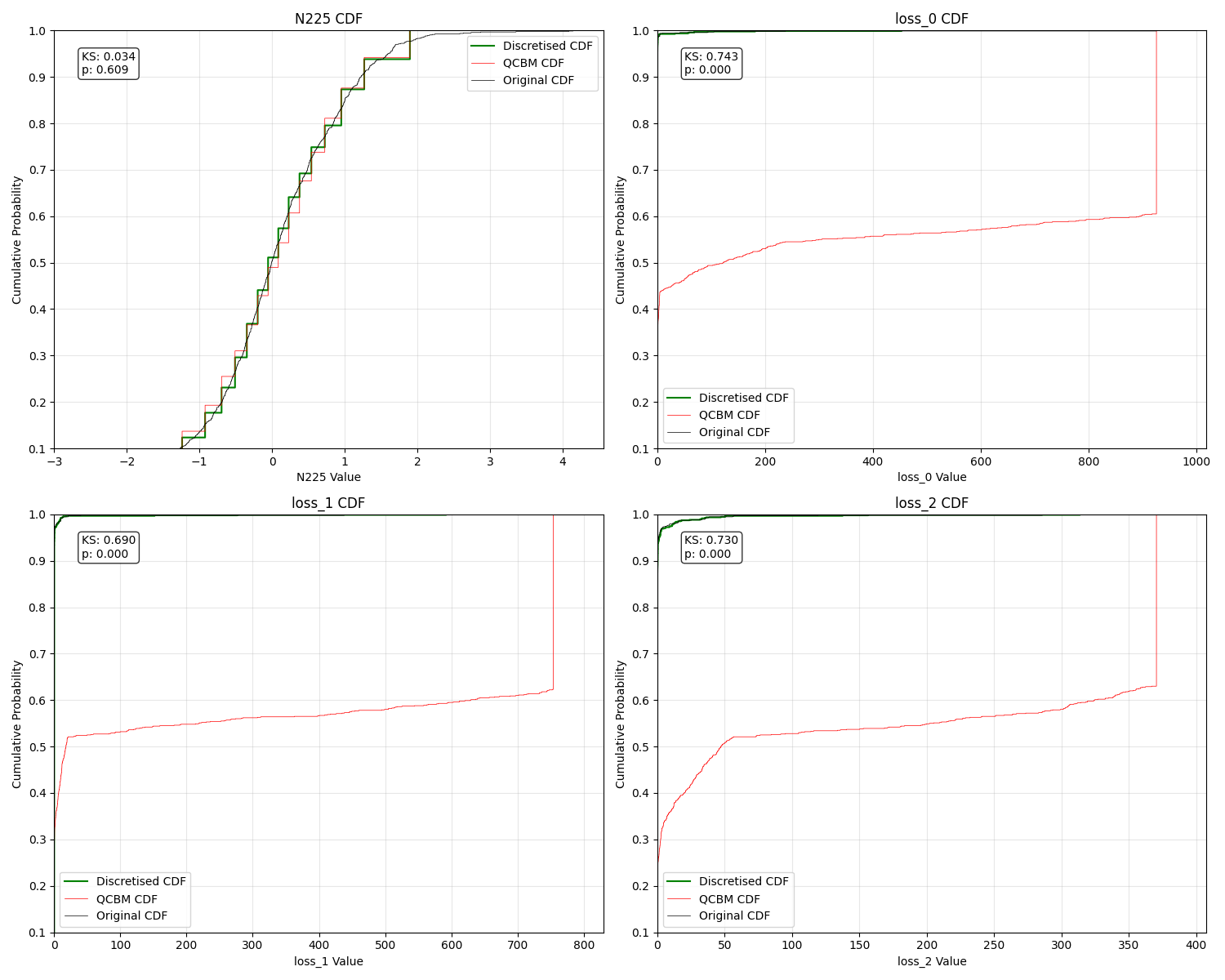}
    \caption{Marginals without hierarchical learning}
    \label{fig:no_hierarchical_learning_aggregate}
  \end{subfigure}
  \caption{Marginal distributions for the 4-qubit-per-variable model with and without hierarchical learning. Hierarchical learning substantially improves the fit of the loss marginals.}
  \label{fig:hierarchical_learning_breakdown}
\end{figure}

Figure~\ref{fig:hierarchical_learning_breakdown} provides additional detail. Without hierarchical learning, the model fits the stock-return marginal more readily than the heavy-tailed loss marginals. This suggests that random initialisation is particularly inadequate for the tail-sensitive part of the problem, whereas hierarchical learning provides a much better starting point for the insurance-loss variables.

These results suggest that hierarchical learning becomes even more important as qubit count and discretisation resolution increase. A more systematic scaling study is left for future work.

\subsection{Effectiveness of exponential encoding}

\begin{figure}
  \centering
  \begin{subfigure}{0.45\textwidth}
    \includegraphics[width=\linewidth]{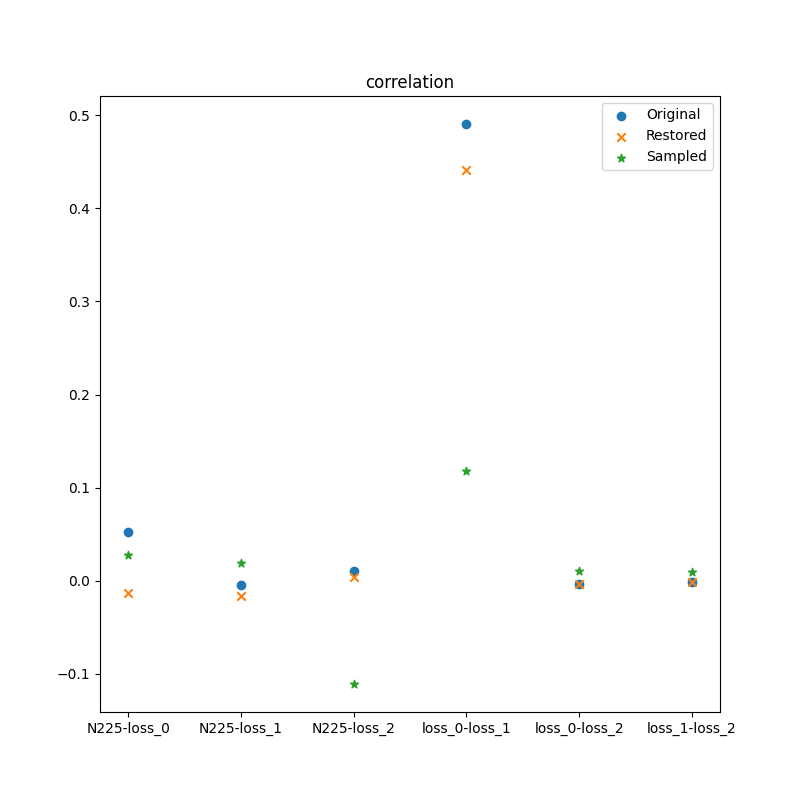}
    \caption{4-qubit model with exponential encoding for losses}
    \label{fig:stockloss_pairwise_corr_qpv=4_exponential}
  \end{subfigure}
  \begin{subfigure}{0.45\textwidth}
    \includegraphics[width=\linewidth]{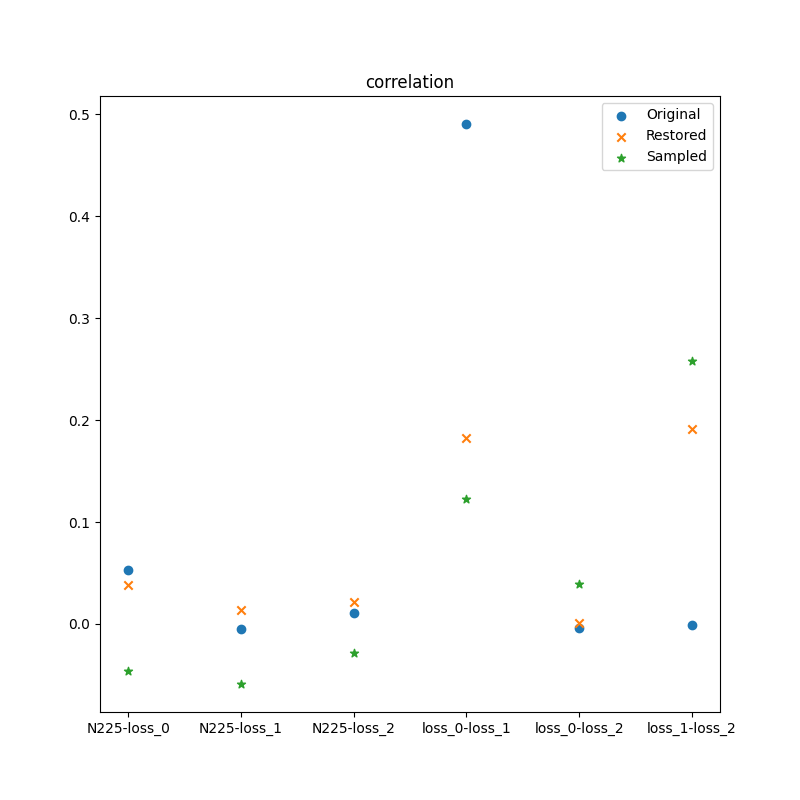}
    \caption{4-qubit model with uniform encoding for losses}
    \label{fig:stockloss_pairwise_corr_qpv=4_uniform}
  \end{subfigure}
  \caption{Pairwise correlations at 4 qubits per variable under exponential and uniform encodings.}
  \label{fig:encoding_qpv4}
\end{figure}

\begin{figure}
  \centering
  \begin{subfigure}{0.45\textwidth}
    \includegraphics[width=\linewidth]{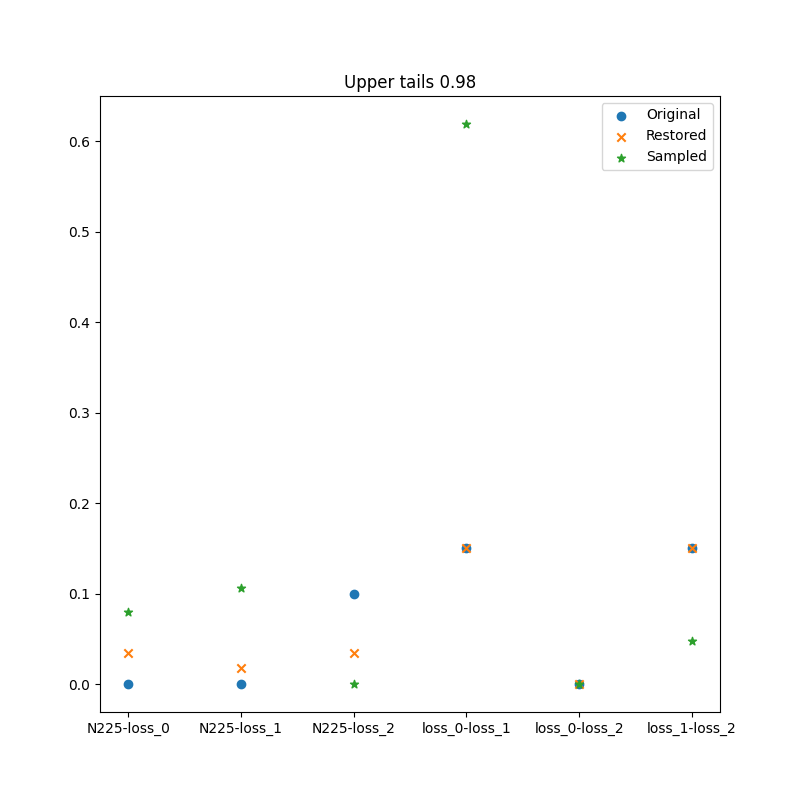}
    \caption{98th-percentile dependence with exponential encoding}
    \label{fig:stockloss_pairwise_corr_p98_qpv=4_exponential}
  \end{subfigure}
  \begin{subfigure}{0.45\textwidth}
    \includegraphics[width=\linewidth]{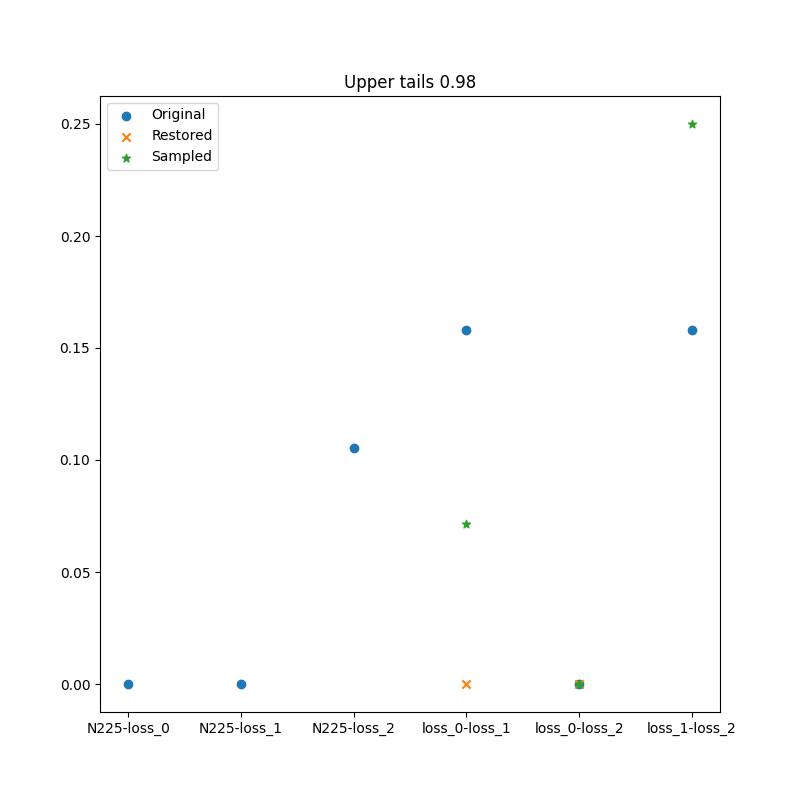}
    \caption{98th-percentile dependence with uniform encoding}
    \label{fig:stockloss_pairwise_corr_p98_qpv=4_uniform}
  \end{subfigure}
  \caption{Pairwise upper-tail co-exceedance at 4 qubits per variable: for each pair, the probability that one variable exceeds its 98th percentile given that the other does.}
  \label{fig:encoding_p98_qpv4}
\end{figure}

Figures~\ref{fig:encoding_qpv4} and \ref{fig:encoding_p98_qpv4} show the recovered pairwise dependence structure under exponential and uniform encodings, both across the full distribution and restricted to the upper tail beyond the 98th percentile.

To quantify tail fidelity, we compare the upper tail of the aggregate portfolio value in the data with samples drawn from each 4-qubit-per-variable model after SPSA (Table~\ref{tab:encoding_tail}). The most important difference is structural. Under the uniform encoding, each basis state is decoded to a single point of the fitted marginal, so the model cannot produce a portfolio value above about 54, and in practice never exceeds 43; the largest observed value in the data is 1366. Under the exponential encoding, the top bin of each loss variable extends to the largest observation, so the model can reach the extreme tail. Neither encoding can generate a value beyond the largest observation of each variable, however, so losses more extreme than any in the data cannot be produced. Consequently, in the extreme tail the exponential model stays within a factor of about two of the data, whereas the uniform model understates it by more than an order of magnitude.

This comes at a cost. Because the top bin is wide and sampled uniformly, the exponential model's upper quantiles are highly variable from one set of samples to the next, and between the 95th and 99.5th percentiles it tends to overshoot the data, where the uniform model is usually closer. For insurance portfolios, the more relevant criterion is the ability to represent extreme losses at all, which the uniform encoding lacks; however, the exponential model's tail estimates should be read together with their sampling variability.

\begin{table}
  \centering
  \caption{Upper tail of the aggregate portfolio value: observed data versus QCBM samples. Model values are medians over 21 sets of 1000 samples from the trained 4-qubit-per-variable models. The uniform encoding cannot exceed about 54, so it misses the extreme tail; the exponential encoding reaches it, but overestimates moderate tail values.}
  \label{tab:encoding_tail}
  \begin{tabular}{lrrr}
    \toprule
     & & \multicolumn{2}{c}{QCBM samples} \\
    \cmidrule(l){3-4}
     & Observed data & Exponential encoding & Uniform encoding \\
    \midrule
    95th percentile & 4.9 & 16.3 & 4.6 \\
    99th percentile & 54.5 & 252.5 & 35.1 \\
    99.9th percentile & 790.8 & 1228.9 & 43.2 \\
    Largest value & 1366.3 & 1552.0 & 43.2 \\
    \bottomrule
  \end{tabular}
\end{table}

\section{Conclusions}
\label{sec:conclusions}

This study shows that QCBMs can be adapted to actuarial copula modelling on insurance-loss data, but only when the training pipeline is designed around the heavy-tailed statistical structure of the problem. Standard QCBM training procedures, developed for lighter-tailed distributions, do not transfer directly.

The main empirical conclusions are as follows:
\begin{enumerate}
    \item \textbf{Exponential quantile encoding} allows the model to represent extreme losses: with a uniform encoding the model cannot produce aggregate values above about 54, against a data maximum of 1366, whereas the exponential encoding reaches the extreme tail, at the cost of overestimating moderate tail values.
    \item \textbf{Randomised KL divergence} provides a stable training signal even when samples are sparse relative to the discretised support, which is what enables the numerical precision of the algorithm to scale to circuit sizes for which the empirical distribution would otherwise be too sparse to inform gradient updates.
    \item \textbf{Hierarchical learning} gives a much better starting point for the larger circuit: in our comparison, it reduced the final randomised KL divergence by a factor of about 2.5 compared with random initialisation followed by SPSA (0.56 vs.\ 1.42).
    \item \textbf{Shot count}, rather than hardware noise, accounts for most of the gap between the noiseless model and H1-1: 1000 shots from an ideal device already give a randomised KL divergence of 1.90, against 2.06 on H1-1 and 1.23 for the exact noiseless model. More shots should narrow this gap, but the number required grows with the number of variables.
\end{enumerate}

More broadly, these results demonstrate an end-to-end pipeline for fitting insurance-portfolio copulas on quantum hardware. Establishing a practical advantage over classical copula methods lies beyond the scope of this work, and we highlight three natural directions for further study: reducing the number of shots needed on hardware, extending the empirical evaluation to larger portfolios and datasets, and carrying out systematic comparisons against classical copula baselines.

\section*{Data Availability Statement}

The data used in this study consist of obfuscated synthetic insurance loss data based on real insurance losses. Due to confidentiality obligations to OXFIDA, the underlying data cannot be deposited in a public repository. Qualified researchers may request access for the purpose of verifying or extending this work by contacting the corresponding author. Requests will be evaluated by OXFIDA on a case-by-case basis and may require the execution of a non-disclosure agreement prior to access being granted.

\section*{Code Availability Statement}

The code used in this study is available from the authors on request.

\backmatter

\begin{appendices}

\section{Experimental details}\label{app:details}

This appendix collects the settings needed to reproduce the experiments. All values were checked against the code and the stored results. Table~\ref{tab:app-data} describes the datasets and their preprocessing, Table~\ref{tab:app-circuit} the circuits, and Table~\ref{tab:app-training} the training settings shared by all experiments. Several settings changed between the October~2025 code that produced the emulator and H1-1 results and the 2026 code that produced the noiseless and hierarchical-learning results; these are listed per experiment in Table~\ref{tab:app-runs}.

\begin{table}[!htbp]
\caption{Datasets and preprocessing.}\label{tab:app-data}
\footnotesize
\begin{tabular*}{\textwidth}{@{\extracolsep{\fill}}>{\raggedright\arraybackslash}p{0.2\textwidth}>{\raggedright\arraybackslash}p{0.42\textwidth}>{\raggedright\arraybackslash}p{0.3\textwidth}@{}}
\toprule
 & Stock/Insurance & Gumbel benchmark \\
\midrule
Source & Obfuscated insurance-loss data (see Data Availability Statement) & \texttt{statsmodels} \texttt{GumbelCopula}, $\theta=5$, $d=4$ \\
Observations & 939 & 2000, regenerated on every run \\
Variables & \texttt{N225}: Nikkei~225 negative returns, standardised to mean~0 and s.d.~1; \texttt{loss\_0}, \texttt{loss\_1}, \texttt{loss\_2}: three anonymised insurance loss lines & four variables with $\mathcal U(0,1)$ margins \\
Share of zero values & \texttt{loss\_0} 86.5\%, \texttt{loss\_1} 76.1\%, \texttt{loss\_2} 84.5\% & none \\
Largest loss & 926.2, 753.9, 370.5 & --- \\
\midrule
Non-loss margins & Best fit among nine \texttt{scipy.stats} families (normal, exponentiated Weibull, Weibull max/min, Pareto, GEV, Student-$t$, gamma, lognormal), chosen by the largest Kolmogorov--Smirnov $p$-value; \texttt{N225} selects Student-$t$ ($\nu\approx7.9$). Probability-integral transform, then $2^n$ equal-width bins on $[0,1]$ & same procedure, applied to each margin \\
Loss margins, exponential encoding & Empirical rank-based bins with boundaries at sorted positions $\lfloor(1-2^{-i})N\rfloor$, $i=0,\dots,2^n-1$, the last bin ending at the maximum; each added uniform bit splits every bin into two bins of equal count. Ties at zero place all zero losses in bin~0, so occupancies are not the nominal 50/25/12.5\% (at 3 qubits, \texttt{loss\_0} occupies bins 0--7 with 812, 0, 10, 59, 29, 15, 7, 7 observations) & not applicable \\
Loss margins, uniform encoding (encoding comparison only) & Point mass at zero mixed with a lognormal fitted to the positive losses; zero maps to bin~0 and the positive part is split into $2^n-1$ equal-probability bins & not applicable \\
Decoding & Uniform draw within the selected bin (exponential) or inverse CDF of a uniform draw within the bin (marginal) & as for \texttt{N225} \\
\midrule
Training data & Full dataset (2026 runs); shuffled 80/20 split, training on 751 rows (October~2025 runs) & 2000 samples; 1600 in October~2025 runs \\
Held-out evaluation & None: all reported comparisons are against the data used for training & none \\
Portfolio aggregate & Unweighted sum of the four columns (standardised \texttt{N225} plus raw losses) & unweighted sum of the four columns \\
\botrule
\end{tabular*}
\end{table}

\begin{table}[!htbp]
\caption{Circuits. Gate counts and depths are for the uncompiled \texttt{pytket} circuits (depth as reported by \texttt{Circuit.depth()}); each $R_{ZZ}$ is implemented as CX--$R_z$--CX.}\label{tab:app-circuit}
\footnotesize
\begin{tabular*}{\textwidth}{@{\extracolsep{\fill}}>{\raggedright\arraybackslash}p{0.27\textwidth}>{\raggedright\arraybackslash}p{0.3\textwidth}>{\raggedright\arraybackslash}p{0.35\textwidth}@{}}
\toprule
 & 3 qubits per variable & 4 qubits per variable \\
\midrule
Qubits & 12 & 16 \\
State preparation (fixed) & \multicolumn{2}{p{0.68\textwidth}}{$H$ on each qubit of the first variable's register, then CNOTs copying each of these qubits to the corresponding qubit of the other three registers} \\
Layers & 2 & 2 \\
Per layer, per variable & $R_z$, $R_x$ on each qubit; ring of 3 $R_{ZZ}$ & $R_z$, $R_x$ on each qubit; ring of 3 $R_{ZZ}$ on the old qubits plus 2 $R_{ZZ}$ linking the new qubit \\
Parameters & 72 & 104 (72 transferred, 32 new) \\
CX gates (incl.\ state preparation) & 57 & 92 \\
Depth (two-qubit depth) & 26 (15) & 38 (23) \\
Angle units & \multicolumn{2}{p{0.68\textwidth}}{half-turns (\texttt{pytket} convention)} \\
\botrule
\end{tabular*}
\end{table}

\begin{table}[!htbp]
\caption{Training and evaluation settings common to all experiments. Settings that differ between experiments are given in Table~\ref{tab:app-runs}.}\label{tab:app-training}
\footnotesize
\begin{tabular*}{\textwidth}{@{\extracolsep{\fill}}>{\raggedright\arraybackslash}p{0.3\textwidth}>{\raggedright\arraybackslash}p{0.64\textwidth}@{}}
\toprule
Setting & Value \\
\midrule
\multicolumn{2}{@{}l}{\emph{Loss}} \\
Objective & Randomised KL divergence $\mathrm{KL}(q_\theta\,\|\,\phi)$ with both arguments clipped below at $\epsilon=10^{-6}$ \\
Model distribution & $q_\theta(x)=|\psi_\theta(x)|/\sum_y|\psi_\theta(y)|$ (Section~\ref{sec:amplitude-objective}) \\
Cuts & $K=10$, drawn once from JAX \texttt{PRNGKey(42)} and kept fixed for the whole run \\
Tail bins in rKL cuts & 2 per variable (the largest and second-largest values) \\
\midrule
\multicolumn{2}{@{}l}{\emph{Stage 1: 3 qubits per variable, state-vector simulator}} \\
Initialisation & Random search; angles drawn uniformly from $[0,1)$ half-turns; the draw with the lowest loss is kept \\
Optimiser & Adam (\texttt{optax}) with global-norm gradient clipping at 1.0 \\
Learning-rate schedule & Staircase exponential decay, factor 0.9 every 10\% of the iteration budget (every 500 steps in the final run) \\
Learning-rate search & Optuna, TPE sampler, learning rate in $(0.04, 0.06)$; 500 iterations per trial, early-stopping patience 400; the Hyperband pruner never triggers because trials report no intermediate values \\
Final run & Up to 5000 iterations, early stopping after 400 iterations without improvement; the best-loss parameters are returned \\
\midrule
\multicolumn{2}{@{}l}{\emph{Stage 2: 4 qubits per variable}} \\
Embedding & 72 trained parameters copied and frozen; 32 new parameters initialised at 0 \\
Optimiser & SPSA (\texttt{qiskit\_algorithms}): learning rate 0.06, perturbation 0.002, 5 resamplings per gradient estimate, blocking with allowed increase 0.01, trust region enabled \\
Returned parameters & Lowest-loss point over all SPSA loss evaluations (the \texttt{last\_avg}=20 average computed by the optimiser is not used) \\
\midrule
\multicolumn{2}{@{}l}{\emph{Hardware and emulator}} \\
Devices & Quantinuum H1-1 and its emulator (default noise model), via Quantinuum Nexus \\
Compilation & Nexus compilation, optimisation level 2 \\
Shots & 1000 per circuit execution \\
Post-processing & Observed frequencies $\hat p$ mapped to $\sqrt{\hat p}/\sum\sqrt{\hat p}$ \\
\midrule
\multicolumn{2}{@{}l}{\emph{Evaluation}} \\
Model samples & 1000 basis states drawn from the model distribution (JAX \texttt{PRNGKey(42)}), decoded as in Table~\ref{tab:app-data} \\
\botrule
\end{tabular*}
\end{table}

\begin{table}[!htbp]
\caption{Settings that differ between experiments. ``Emulator/H1-1'' covers the emulator and H1-1 panels of Figures~\ref{fig:stockloss_results} and~\ref{fig:gumbel_results}; ``Hierarchical learning'' the hierarchical-learning comparison in Section~\ref{sec:results}; ``Noiseless and encoding'' the noiseless 4-qubit Stock/Insurance results and the encoding comparison. The random-initialisation arm of the hierarchical-learning comparison uses 4 exponential bits per variable (96 parameters, none transferred), the best of 1000 random draws, then 300 SPSA iterations on all parameters. The Stock/Insurance emulator and H1-1 results share identical trained weights and differ only in the final evaluation; the Gumbel emulator and H1-1 results come from separately trained parameters.}\label{tab:app-runs}
\footnotesize
\begin{tabular*}{\textwidth}{@{\extracolsep{\fill}}>{\raggedright\arraybackslash}p{0.2\textwidth}>{\raggedright\arraybackslash}p{0.25\textwidth}>{\raggedright\arraybackslash}p{0.22\textwidth}>{\raggedright\arraybackslash}p{0.23\textwidth}@{}}
\toprule
 & Emulator/H1-1 & Hierarchical learning & Noiseless and encoding \\
\midrule
Date of runs & October 2025 & March 2026 & April--June 2026 \\
Training data & 80/20 split & full dataset & full dataset \\
New-qubit $R_{ZZ}$ gates & both act on $(q_2,q_0)$, so $q_3$ is not entangled within the layer & $(q_2,q_3)$ and $(q_3,q_0)$ & $(q_2,q_3)$ and $(q_3,q_0)$ \\
Initial random search & 100 (Stock/Insurance, Gumbel emulator); 30 (Gumbel H1-1) & 1000 & 900 \\
Optuna trials & 5 & 30 & 30 (learning rate 0.0406) \\
Seeds (NumPy / Optuna) & 42 / 42 & 46 / 46 & 46 / 46 \\
SPSA iterations & 100; 1 for Gumbel H1-1 & 300 per arm & 500 (noiseless 4-qubit; exponential in encoding comparison); 300 (uniform in encoding comparison) \\
SPSA loss evaluated on & H1 emulator, 1000 shots per evaluation & state vector & state vector \\
Final evaluation & One 1000-shot execution on the emulator or H1-1 & state vector & state vector \\
\botrule
\end{tabular*}
\end{table}

\clearpage

\end{appendices}

\end{document}